\documentclass{wscpaperproc}
\usepackage{latexsym}
\usepackage{caption}
\usepackage{subcaption}
\usepackage{graphicx}
\usepackage{mathptmx}
\usepackage{lipsum}
\usepackage{booktabs}
\usepackage{empheq,amsmath}
\usepackage{amsfonts}
\usepackage{amssymb}
\usepackage{amsbsy}
\usepackage{amsthm}
\usepackage{commath}
\usepackage{nicefrac}
\usepackage{float}
\usepackage{mathtools}
\usepackage{enumitem}
\usepackage{multicol}
\setlist{nosep}

\usepackage{cuted}

\newcommand{\citeauthor}[1]{\citeauthoryear{#1}}

\theoremstyle{plain}

\newtheorem{obsv}{Observation}

\usepackage[pdftex,colorlinks=true,urlcolor=blue,citecolor=black,anchorcolor=black,linkcolor=black]{hyperref}

\newtheoremstyle{wsc}
{3pt}
{3pt}
{}
{}
{\bf}
{}
{.5em}
{}

\theoremstyle{wsc}

\begin{document}

%
%

\pagestyle{fancyplain}

\thispagestyle{plain}
\firstPageHead{}

\chead{\fancyplain{}{\itshape Shen, Achar and Lopes}}

\rhead{}
\cfoot{}
\renewcommand{\headrulewidth}{0pt} 

\makeatletter
\let\@internalcite\cite
\def\cite{\def\@citeseppen{-1000}%
    \def\@cite##1##2{(##1\if@tempswa , ##2\fi)}%
    \def\citeauthoryear##1##2##3{##1 ##3}\@internalcite}
\def\citeNP{\def\@citeseppen{-1000}%
    \def\@cite##1##2{##1\if@tempswa , ##2\fi}%
    \def\citeauthoryear##1##2##3{##1 ##3}\@internalcite}
\def\citeN{\def\@citeseppen{-1000}%
    \def\@cite##1##2{##1\if@tempswa, ##2)\else{}\fi}%
    \def\citeauthoryear##1##2##3{##1 (##3)}\@citedata}
\def\citeA{\def\@citeseppen{-1000}%
    \def\@cite##1##2{(##1\if@tempswa , ##2\fi)}%
    \def\citeauthoryear##1##2##3{##1}\@internalcite}
\def\citeANP{\def\@citeseppen{-1000}%
    \def\@cite##1##2{##1\if@tempswa , ##2\fi}%
    \def\citeauthoryear##1##2##3{##1}\@internalcite}
\def\shortcite{\def\@citeseppen{-1000}%
    \def\@cite##1##2{(##1\if@tempswa , ##2\fi)}%
    \def\citeauthoryear##1##2##3{##2 ##3}\@internalcite}
\def\shortciteNP{\def\@citeseppen{-1000}%
    \def\@cite##1##2{##1\if@tempswa , ##2\fi}%
    \def\citeauthoryear##1##2##3{##2 ##3}\@internalcite}
\def\shortciteN{\def\@citeseppen{-1000}%
    \def\@cite##1##2{##1\if@tempswa, ##2\else{}\fi}%
    \def\citeauthoryear##1##2##3{##2 (##3)}\@citedata}
\def\shortciteA{\def\@citeseppen{-1000}%
    \def\@cite##1##2{(##1\if@tempswa , ##2\fi)}%
    \def\citeauthoryear##1##2##3{##2}\@internalcite}
\def\shortciteANP{\def\@citeseppen{-1000}%
    \def\@cite##1##2{##1\if@tempswa , ##2\fi}%
    \def\citeauthoryear##1##2##3{##2}\@internalcite}
\def\citeyear{\def\@citeseppen{-1000}%
    \def\@cite##1##2{(##1\if@tempswa , ##2\fi)}%
    \def\citeauthoryear##1##2##3{##3}\@citedata}
\def\citeyearNP{\def\@citeseppen{-1000}%
    \def\@cite##1##2{##1\if@tempswa , ##2\fi}%
    \def\citeauthoryear##1##2##3{##3}\@citedata}
%
%
%
\def\@citedata{%
    \@ifnextchar [{\@tempswatrue\@citedatax}%
                  {\@tempswafalse\@citedatax[]}%
}

\def\@citedatax[#1]#2{%
\if@filesw\immediate\write\@auxout{\string\citation{#2}}\fi%
  \def\@citea{}\@cite{\@for\@citeb:=#2\do%
    {\@citea\def\@citea{, }\@ifundefined
       {b@\@citeb}{{\bf ?}%
       \@warning{Citation `\@citeb' on page \thepage \space undefined}}%
{\csname b@\@citeb\endcsname}}}{#1}}%

%
\def\@citex[#1]#2{%
\if@filesw\immediate\write\@auxout{\string\citation{#2}}\fi%
  \def\@citea{}\@cite{\@for\@citeb:=#2\do%
    {\@citea\def\@citea{; }\@ifundefined
       {b@\@citeb}{{\bf ?}%
       \@warning{Citation `\@citeb' on page \thepage \space undefined}}%
{\csname b@\@citeb\endcsname}}}{#1}}%

%
\def\@biblabel#1{}
\makeatother



\newdimen\bibindent
\bibindent=0.0em
\def\thebibliography#1{\section*{\refname}\list
   {}{\settowidth\labelwidth{[#1]}
   \leftmargin\parindent
   \itemindent -\parindent
   \listparindent \itemindent
   \itemsep 0pt
   \parsep 0pt}
   \def\newblock{}
   \sloppy
   \sfcode`\.=1000\relax}


\setlength{\baselineskip}{12.7pt}

\title{\uppercase{Toward Understanding the Impact of User Participation in Autonomous Ridesharing Systems}}


\author{\bf Wen  Shen, Rohan Achar, and  Cristina V. Lopes\\
University of California, Irvine \\
Irvine, CA 92697, USA \\
\{wen.shen, rachar, lopes\}@uci.edu
}

\maketitle

\section*{ABSTRACT}
Autonomous ridesharing systems (ARS) promise many societal and environmental benefits,  including decreased accident rates, reduced energy consumption and pollutant emissions, and diminished land use for parking. To unleash ARS' potential, stakeholders must understand how the degree of passenger participation influences the ridesharing systems' efficiency. To date, however, a careful study that quantifies the impact of user participation on ARS' performance is missing. Here, we present the first simulation analysis to investigate how and to what extent user participation affects the efficiency of ARS. We demonstrate how specific configurations (e.g., fleet size, vehicle capacity, and the maximum waiting time) of a system can be identified to counter the performance loss due to users' uncoordinated behavior on ridesharing participation.  Our results indicate that stakeholders  of ARS should base decisions regarding system configurations on insights from data-driven simulations and make tradeoffs between system efficiency and price of anarchy for desired outcomes.


\section{\uppercase{Introduction}}
Real-time ridesharing  (or dynamic carpooling) services have emerged as a convenient and cost-efficient option for commuters to mobilize themselves in cities~\shortcite{furuhata2013ridesharing}. These ridesharing systems are transforming urban transportation  in an economically feasible,  environment-friendly, and socially beneficial way~\shortcite{furuhata2013ridesharing,shen2016online}. Recent advances in autonomous driving technology accelerate this transformation  because  they have the potential to evolve traditional carpooling services into {\em autonomous ridesharing systems} ({\bf ARS})~\shortcite{santi2014quantifying,shen2015managing,shen2016online,alonso2017demand}. Through system-wide coordination,  ARS  promise a host of social and environmental benefits, including alleviated traffic congestion, reduced energy consumption and greenhouse gas emissions, less need for land use, as well as increased safety~\shortcite{shen2015managing,shen2016online}. 

%

Prior studies show that three key factors influence the performance of autonomous ridesharing systems~\shortcite{santi2014quantifying,gargiulo2015dynamic,shen2016online,alonso2017demand}.  The three factors are: (1) the trip-vehicle assignment (i.e., scheduling, routing, and pricing) algorithms used to match the requests to the vehicles and to compute the fares for the trips~\shortcite{cordeau2006branch,santi2014quantifying,shen2016online,alonso2017demand};  (2) the physical structure of the transportation network, such as road networks, the fleet size,  the vehicle capacity and the request distribution~\shortcite{alonso2017demand};  and (3) user participation~\shortcite{shen2015managing,shen2016online,gargiulo2015dynamic}. To unleash ARS' potential, it is important for the stakeholders to understand how these factors influence the systems' performance~\shortcite{shen2015managing,gargiulo2015dynamic,alonso2017demand}.

Despite much research on ridesharing, most of them focus on the dynamic assignment problem and the network infrastructure~\shortcite{cordeau2006branch,santos2013dynamic,shen2016online,santi2014quantifying,alonso2017demand}. On the one hand, stakeholders of ARS might not be able to implement efficient policies to regulate the systems for societal benefits without knowing how users' individual behavior influences the systems' efficiency~\shortcite{shen2017regulating}. 
On the other hand, little attention has been paid to understanding how passengers' uncoordinated behavior on deciding whether to opt in ridesharing or not affects the systems' performance (e.g., total operational cost).

To fill the gap,  we present the first simulation analysis to quantify the impact of user participation on the efficiency of autonomous ridesharing systems with different configurations. Our work advances the state of the art in the following ways:
\begin{enumerate}
\item[$\bullet$] We describe a modular, agent-based platform, called {\em SpaceTime for Autonomous Ridesharing Systems} ({\bf STARS}),  for simulating large-scale autonomous ridesharing systems. 
\item[$\bullet$] We conducted extensive experiments on the STARS platform with 96 different sets of system configurations using the data extracted from the New York City Taxi trip public data. Experimental results demonstrate that tradeoffs between system efficiency and the price of anarchy are necessary and are typically viable in order to achieve desired outcomes.

\end{enumerate}

\section{\uppercase{Simulating Autonomous Ridesharing Systems}}
This section discusses methods for solving the trip-vehicle assignment problem, and approaches for modeling users' choice of ridesharing participation.

\subsection{Trip-Vehicle Assignment}

Traditional algorithms (e.g., the branch-and-cut algorithm~\shortcite{cordeau2006branch},  and the annealing meta-heuristic algorithm~\shortcite{braekers2014exact})  for the trip-vehicle assignment problem usually suffer scalability problems and are not applicable to large ridesharing systems that have thousands of vehicles, requests, and streets. This is primarily due to the large search space and the spatial-temporal constraints in the dynamic ridesharing problem~\shortcite{furuhata2013ridesharing,alonso2017demand}. 
%


Recently, \shortciteN{alonso2017demand} extended the static shareability network model by~\shortciteN{santi2014quantifying} to dynamic ridesharing with high-capacity vehicles (serving up to ten passengers at the same time). They introduced an anytime algorithm that can compute optimal solutions to the trip-vehicle assignment for large-scale ridesharing systems in real time~\shortcite{alonso2017demand}. The method consists of the following three steps: generating the {\em request-vehicle} ({\bf RV}) graph, generating the {\em request-trip-vehicle} ({\bf RTV}) graph, and computing the optimal assignment. The anytime approach also runs a load rebalancing process to redistribute the idle vehicles to pick up unassigned requests.

The STARS  platform in our work utilizes the anytime algorithm to allow simulating real-time operations of city-scale ridesharing systems. 
%
\subsection{Modeling User Participation}
 In our work, we used {\em Price of Anarchy} ({\bf PoA})~\shortcite{roughgarden2005selfish} as the indicator to quantify the extent to which  passengers' uncoordinated  behavior of ridesharing participation  diminishes  ARS' performance. The is because price of anarchy has been used as a measure to quantify how the efficiency of a system degrades due to uncoordinated behavior of its agents in many areas, including transportation systems~\shortcite{youn2008price}, smart buildings and power grids~\shortcite{shen2013regulating,shen2017regulating}. 
 
 We take a data-driven approach to compute the price of anarchy. Specifically, we select different portions of the passengers as participants, and the rest of the population as non-participating riders. We assume that each of the selected choice profiles follows a stochastic user equilibrium~\shortcite{daganzo1977stochastic}: no user believes she can improve her utility by unilaterally changing her decision on whether to participate to not. This assumption allows us to model passengers' choice behavior based on real-world data instances rather than synthesized samples. Similar assumptions are also made in literature~\shortcite{youn2008price}.

\section{\uppercase{Simulation Platform}}
In this section, we describe the {\em SpaceTime for Autonomous Ridesharing Systems} ({\bf STARS}) simulation platform. We first introduce the basic design of STARS and then discuss details of its implementation. 

\subsection{System Design}

\begin{figure}[h]
\centering
\begin{tabular}{rl}
  \begin{subfigure}[b]{0.45\linewidth}
  \includegraphics[width=0.98\linewidth]{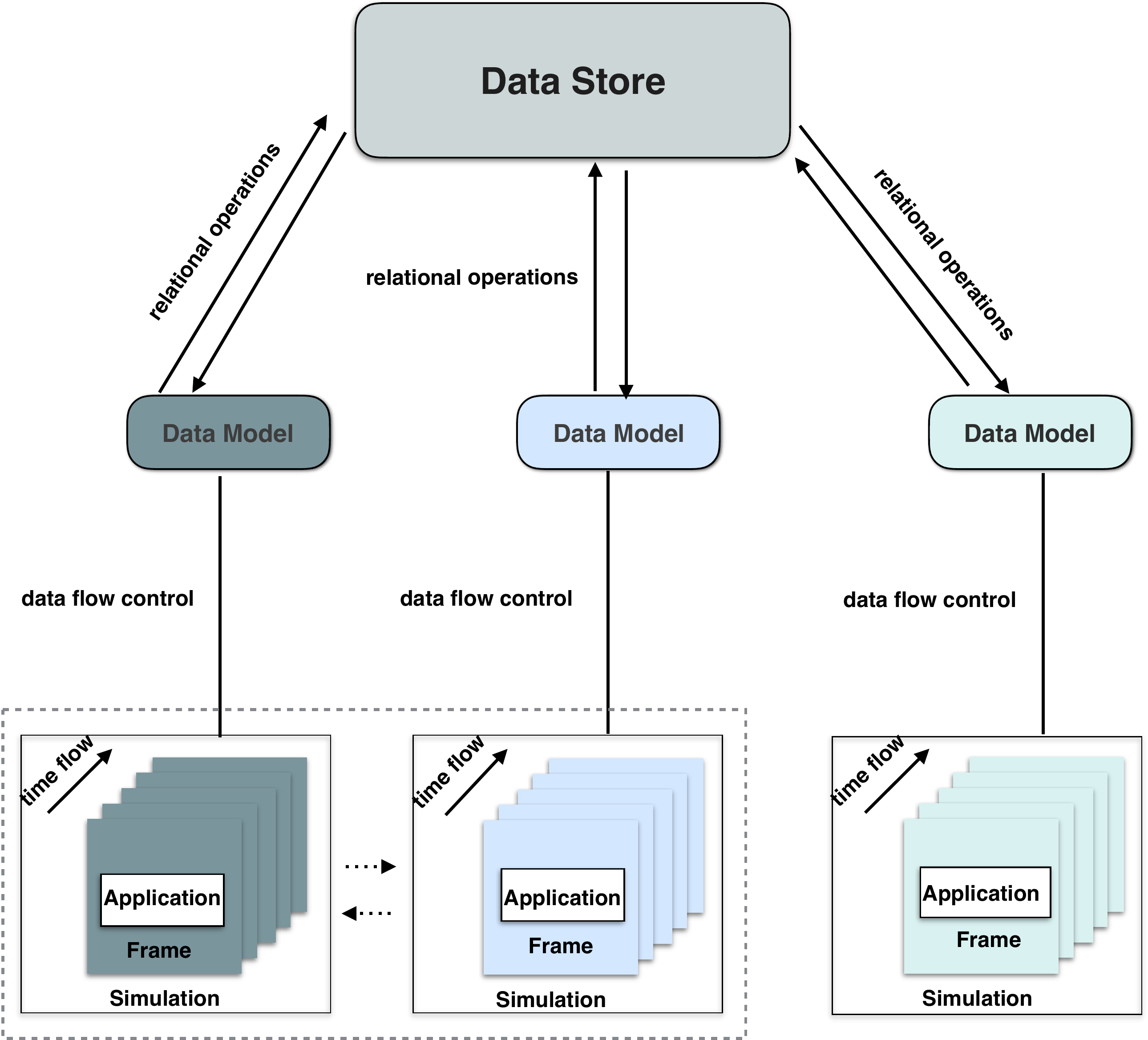}
  \caption{system design}
     \label{fig:spacetime}
\end{subfigure}\hfill
 
\begin{subfigure}[b]{0.45\linewidth}
\includegraphics[width=.98\linewidth]{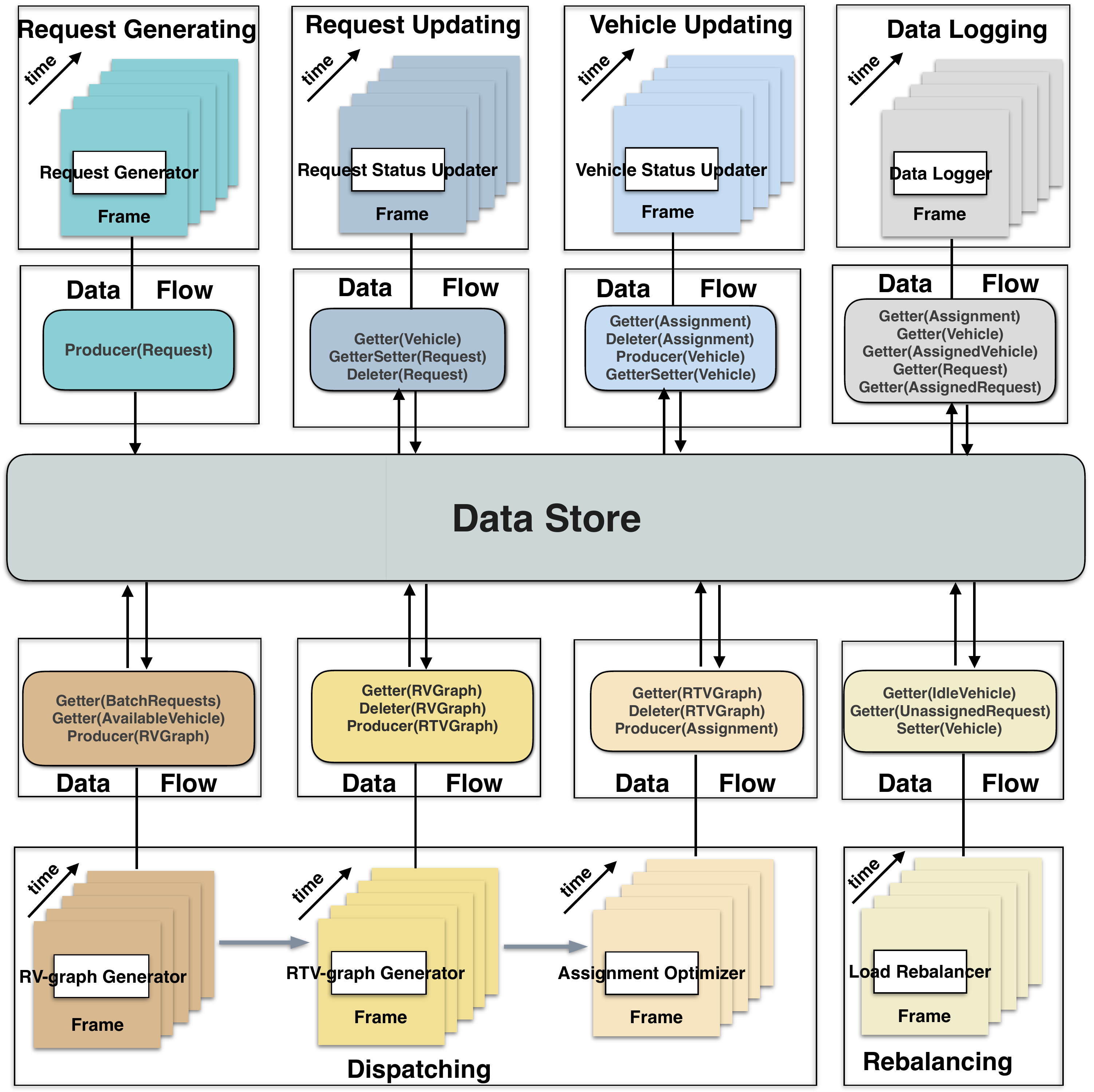}
\caption{architectural view}
     \label{fig:ridesharing}
\end{subfigure}
 \end{tabular}
 \caption{The STARS simulation platform.}
\label{fig:stdr}
\end{figure}

STARS is a multi-agent simulation platform  on top of a simulation framework called {\em Spacetime}~\shortcite{valadares2016cadis,lopes2017predicate}.  Spacetime follows a proof-of-concept paradigm called {\em Predicate Collection Classes} ({\bf PCCs})~\shortcite{lopes2017predicate} to increase modularity.  The design of STARS consists of five aspects: {\em frame}, {\em time flow}, {\em data flow control}, {\em data model}, and {\em data store} (See Figure~\ref{fig:spacetime} for an abstract view of STARS). We discuss each in turn as follows:
\begin{enumerate}
\item[$\bullet$]{\bf Frame} allows system designers to attach applications, fetch objects of desired types, add new objects and delete objects every time step. 
\item[$\bullet$]{\bf Data flow control} declares each application's interaction with its PCCs. There are five main declarations:
\begin{enumerate}
\item[---] {\em Producer: }the application creates objects of specified types;
\item[---] {\em Getter:} the application fetches objects of specified types; 
\item[---] {\em Setter:} the application updates objects of specified types;
\item[---] {\em GetterSetter:} the application fetches and updates objects of specified types; 
\item[---] {\em Deleter:} the application deletes object of specified types.
\end{enumerate}
\item[$\bullet$] {\bf Time flow} defines the execution frequency of loop operations (i.e., {\em pull}, {\em update} and {\em push}). The {\em pull} operation retrieves from the data store to get current information of the objects whose types are included in {\em Getter} and {\em GetterSetter}. The {\em update} operation calls the update function during which the {\em Frame} collects all the modifications of the setter objects whose types are included in  {\em GetterSetter}. In the {\em push} operation,  the {\em Frame} send all the modifications collected in the {\em update} operation to the data store.   The loop operations are performed every fixed interval of clock time. 
\item[$\bullet$] {\bf Data model} defines the objects of the simulation in terms of classes that are both types and sets. These classes are called {\em Predicate Collection Classes (PCCs)}~\shortcite{lopes2017predicate}.
\item[$\bullet$] {\bf Data store} keeps track of the PCC objects. It assumes a distributed simulation environment using pass-by-copy to fulfill data requests,  and handles the operation of pushing state updates back.
\end{enumerate}

In the original design of {Spacetime}, each application is encapsulated and is independent of each other~\shortcite{valadares2016cadis,lopes2017predicate}. However, multi-agent simulations typically require interactions between different applications (agents)~\shortcite{macal2010tutorial}.  Different from {Spacetime},  STARS enables communications between different applications on top of each individual application agent. The communications between the interacting applications also follow a chronological order (i.e., arranged in order by time).  With a dedicated modular design, the STARS platform supports complex, scalable simulations in a distributed, or multicore environment, making it a good fit for autonomous ridesharing simulations. 
\subsection{Platform Implementation}
We next report the  implementation details of the STARS platform.  STARS comprises five simulation agents: the {\em request generating agent} that reads requests from raw data, the {\em request updating agent} that refreshes the status of requests, the {\em vehicle updating agent} that initializes vehicles and updates vehicle status, the {\em data logging agent} that records the state of the ridesharing sytem, the {\em dispatching agent} that computes optimal assignments, and the {\em load rebalancing agent} that redistributes vehicles for picking up unassigned requests. Among them, the {\em dispatching agent} is composed of three working agents: the RV-graph generator, the RTV-graph generator and the assignment optimizer.  The RV-graph generator computes which requests can be paired and which vehicles can service which requests individually.  The RTV-graph generator calculates feasible trips that can be combined and picked up by a vehicle.  The assignment optimizer computes the optimal assignment using the anytime algorithm described in the work by~\shortciteN{alonso2017demand}.

In STARS, each agent independently fulfills its role in the simulation within a frame that has predefined computational space and time flow. In what follows we discuss the role of each agent in turn.

\begin{enumerate}
\item[$\bullet$] {\bf Request Generating: } The request generating agent produces request objects and pushes them to the data store every time interval $\theta$ ($\theta=1s$ in our experiments). A request is defined  as a tuple of origin, destination, the time of the request, the latest acceptable pick-up time, the pick-up time, the expected drop off time, and the earliest possible time for reaching the destination. 
\item[$\bullet$] {\bf Request Updating: } Given the locations of the vehicles and the vehicles' passengers (the requests that are being serviced by the vehicles), the request updating agent refreshes the status of the requests  every time interval $\theta$ ($\theta=0.5s$ in our experiments). If the destination of a request has been reached, the request will be deleted.
\item[$\bullet$] {\bf Data Logging: } The agent records the state of the assignment, vehicles, and the requests every time interval $\theta$ ($\theta=0.5s$ in our experiments).
\item[$\bullet$] {\bf Dispatching: } The dispatching agent performs three steps using the algorithm described in~\shortcite{alonso2017demand}: Computing the pairwise request-vehicle RV graph, generating the request-trip-vehicle RTV graph and calculating the optimal assignment. We elaborate each step in turn as follows:
\begin{enumerate}
\item[---] {\em Computing RV graph:} Given available vehicles and a batch of requests, this step is to compute the requests that can be pairwise combined, and to search the vehicles that can serve which requests individually ($\theta = 30s$ in our experiments).
\item[---] {\em Generating RTV graph:}  Given the RV graph, this step is to find feasible trips that can be combined and picked up by a vehicle subject to all the constraints. It is done by computing complete subgraphs or cliques of the RV graph ($\theta = 30s$ in our experiments).
\item[---] {\em Calculating the optimal assignment:} Given the RTV graph, this step is to calculate the optimal assignment of vehicles to trips with one trip per vehicle at most. It is solved incrementally by formalizing the problem into an  Integer Linear Program ($\theta = 30s$ in our experiments).
\end{enumerate}
\item[$\bullet$] {\bf Vehicle Updating: } Given the assignment computed by the dispatching agent, the vehicle updating agent adds the requests in the assigned trip to the respective vehicles. Once the an assignment has been processed, it will be deleted from the data store.  The agent updates the location of each vehicle according to the designated route associated with it ($\theta = 0.5s$ in our experiments). The vehicles are initialized at the beginning of the simulation.
\item[$\bullet$] {\bf Load Rebalancing: } Given the idle vehicles, and the unassigned requests, the rebalancing agent redistributes vehicles to the locations of unassigned requests ($\theta = 30s$ in our experiments). Specifically, after each assignment, if there are some unassigned requests and some available vehicles, the rebalancing process is executed. 
\end{enumerate}

STARS adopts a modular design, making it easily expandable and configurable.  By using the state-of-the-art dispatching algorithm~\shortcite{alonso2017demand}, STARS is capable of simulating real-time operations of autonomous ridesharing systems that have thousands of requests and vehicles at the same time in large road networks.  Thus, it is feasible to use STARS to conduct extensive simulations with real-world data in ridesharing to gain insights on how user participation influences the efficiency of autonomous ridesharing systems.

\section{\uppercase{Data  and Methods}}
This section describes the experimental settings, including the dataset used in the simulations, and the procedures followed to investigate the impact of user participation in autonomous ridesharing systems.
\subsection{Dataset}
We used the New York City Taxi trip Dataset~\cite{nyctaxidata} in the experiments. We selected a typical week (i.e., a week without major holidays included) in the year of 2011, from Monday, June 6th 00:00:00,  to Sunday, June 12th 23:59:59, as the time duration for simulating the operation of the autonomous ridesharing systems. We extracted the whole week's trips of which the origins and the destinations were within Manhattan from the raw dataset. The resulting dataset contains 3,014,628 trips ranging from 391,246 (Saturday) to 465,331(Monday) per day. The trips were originally serviced by a fleet of 13,586 taxis. Each trip in the raw data has the following attributes: pickup datetime, dropoff datetime, passenger count, pickup longitude, pickup latitude, dropoff longitude, and dropoff latitude. 

In this work, we defined a request $r$ as a tuple $(o_{r}, d_{r}, tr_{r}, tlp_{r}, tp_{r}$, $ td_{r}, te_{r},  m_{r})$, where $o_{r}$ denotes the origin, $d_{r}$ is the destination, $tr_{r}$ represents the time of the request, $tlp_{r}$ is the latest acceptable pickup time, $tp_{r}$ denotes the pickup time, $td_{r}$ is the expected dropoff time, $te_{r}$ denotes the earliest dropoff time possible, and  $m_{r}$ indicates user participation -- user $r$'s decision on whether to participate in a shared ride (i.e., $m_{r} = 1$) or to opt for a private one (i.e., $m_{r} = 0$). Since the raw data does not include all the information as required for autonomous ridesharing simulations, we processed the data with the following procedures: we used the pickup datetime as the request time. We calculated the latest acceptable pickup time by $tlp_{r} = tr_{r} + \Omega$, where $\Omega$ is the maximum waiting time allowed in a ridesharing system. The earliest dropoff time possible $te_{r}$ is computed by $te_{r} = tr_{r} + \tau (o_{r}, d_{r})$, 
where $\tau(o_{r}, d_{r})$ is the shortest travel time from $o_{r}$ to $d_{r}$. The total delay $\delta_{r}$ due to ridesharing for a serviced request $r$ is calculated by  $\delta_{r} = td_{r}- te_{r}$.
In our experiments, we required that $\delta_{r} \leq \Delta$ for all serviced requests, where $\Delta$ is the maximum delay allowed in a ridesharing system. That is, for all $r$, we have $td_{r} \leq te_{r} + \Delta$.

We converted the road map of Manhattan into a graph with 4,092 nodes and 9,453 edges. We then estimated the hourly travel time on each edge for each day of the week by using the method described in the work by~\shortciteN{santi2014quantifying}. With the road network and the travel time estimate, we precomputed the shortest paths and travel time between all the nodes in the network and then stored the calculated results in a lookup table for later use.
%

\subsection{Methods}
To study how user participation affects ARS'  performance, we selected the following metrics: the service rate (the percentage of requests serviced, the higher the better), the mean waiting time (the smaller the better), the mean in-car delay (the difference between the travel delay and the waiting time, the smaller the better), the percentage of shared rides (the higher the better), the mean travel distance (the smaller the better) and the total cost (the smaller the better).  Let $R$ denote the set of requests, $P$ represent the set of profiles that characterize all the passengers' choices on participation, and $p = (m_{r})_{r\in R} $ be a generic profile in $P$. The total cost $\mathcal{C}(p)$ of the system under the profile $p$ is the sum of travel delay $\delta_{r}$ of all serviced requests $r \in R_{o} \subseteq R$ plus a constant cost $c_{d}$ ($c_{d}=1000$ in our experiments) for each denied request $r \in R_{d} = R\setminus R_{o}$. That is, 
\begin{equation}
\mathcal{C}(p) = \sum_{r \in R_{o}} \delta_{r} + \sum_{r \in R_{d}} c_{d} \; .
\end{equation}
The price of anarchy due to passengers' uncoordinated  behavior of ridesharing participation  is defined as
\begin{equation}
PoA = \frac{\max_{p\in P_{sue}}\mathcal{C}(p)}{\min_{p \in P}\mathcal{C}(p)} \;,
\end{equation}
where $P_{sue} \subseteq P$ is the set of choice profiles under stochastic user equilibria.

\begin{figure}[h]
\centering
\begin{tabular}{ccc}
  \begin{subfigure}[b]{0.30\linewidth}
  \includegraphics[width=0.93\linewidth]{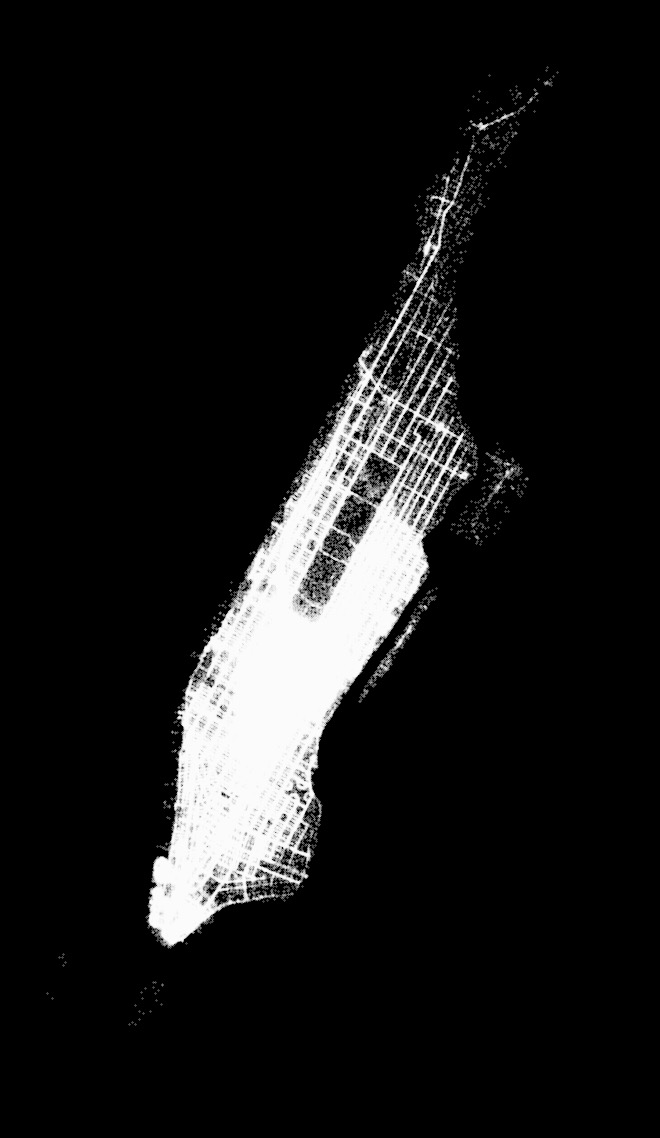}
  \caption{Request distribution}
     \label{fig:reqdistr}
\end{subfigure}\hfill
 
\begin{subfigure}[b]{0.30\linewidth}
\includegraphics[width=.91\linewidth]{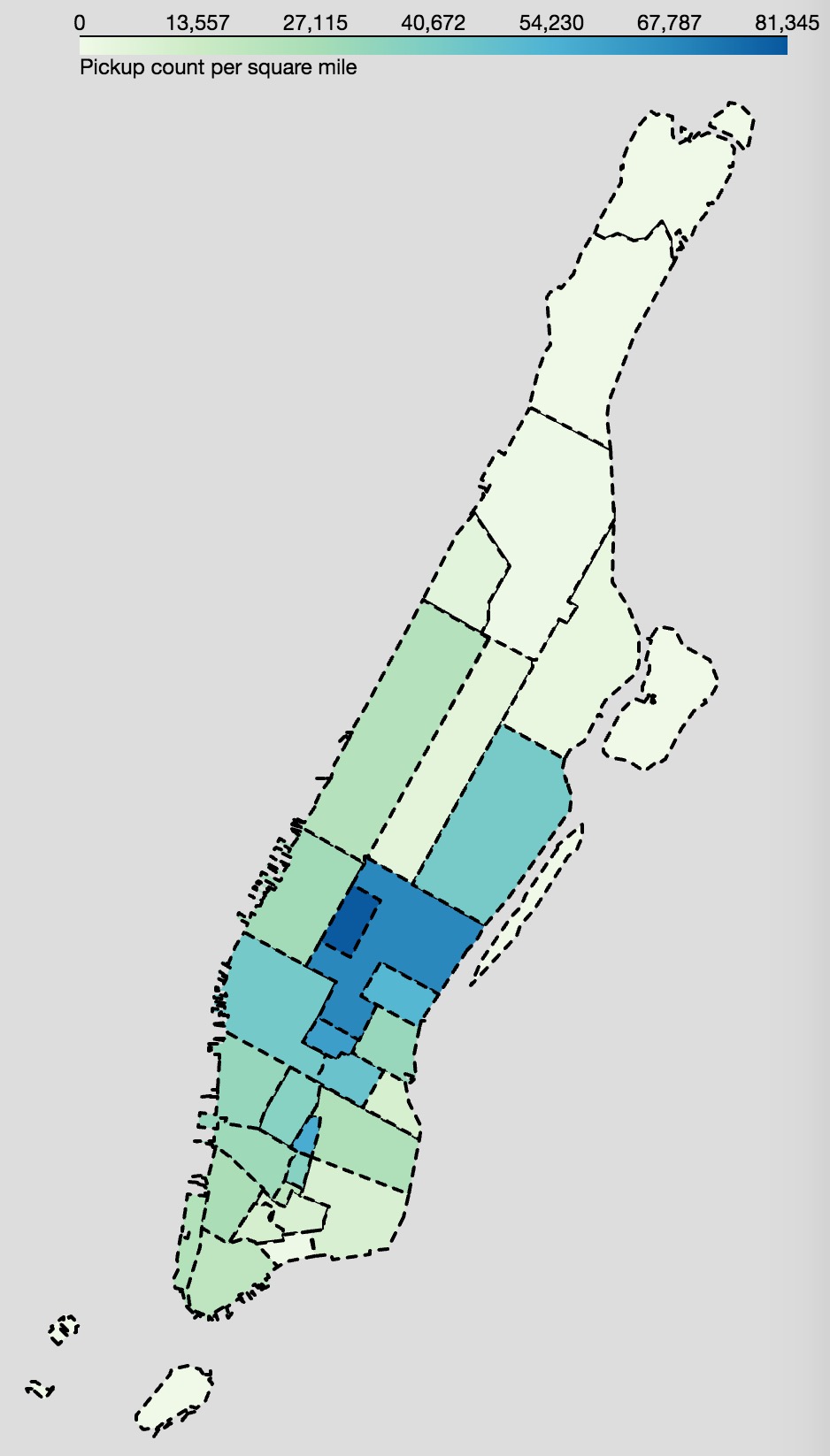}
\caption{Request density}
     \label{fig:neighborhooddensity}
\end{subfigure}

\begin{subfigure}[b]{0.3\linewidth}
\includegraphics[width=\linewidth]{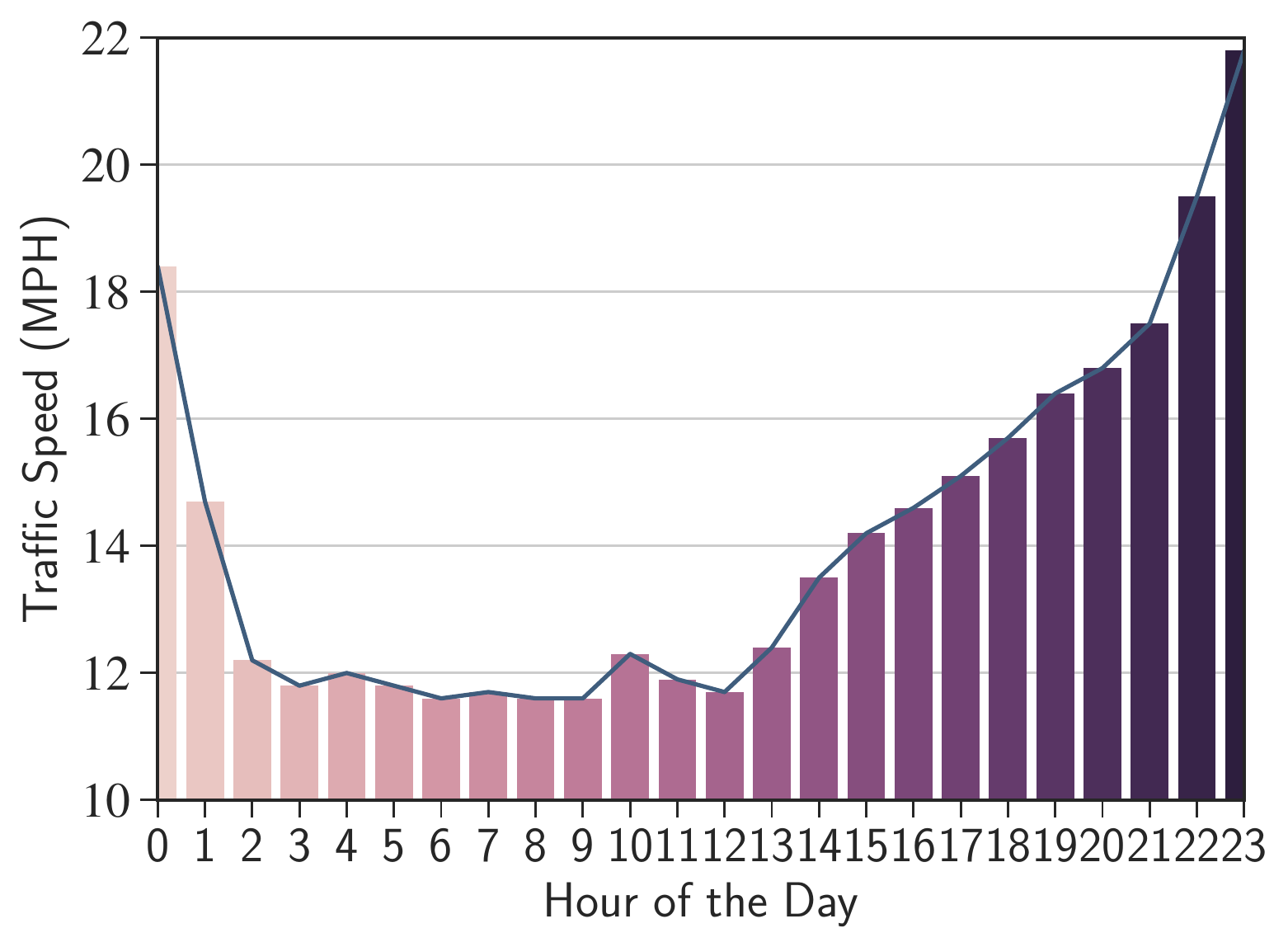}
\caption{Traffic speed}
     \label{fig:hourspeed}
    
\includegraphics[width=\linewidth]{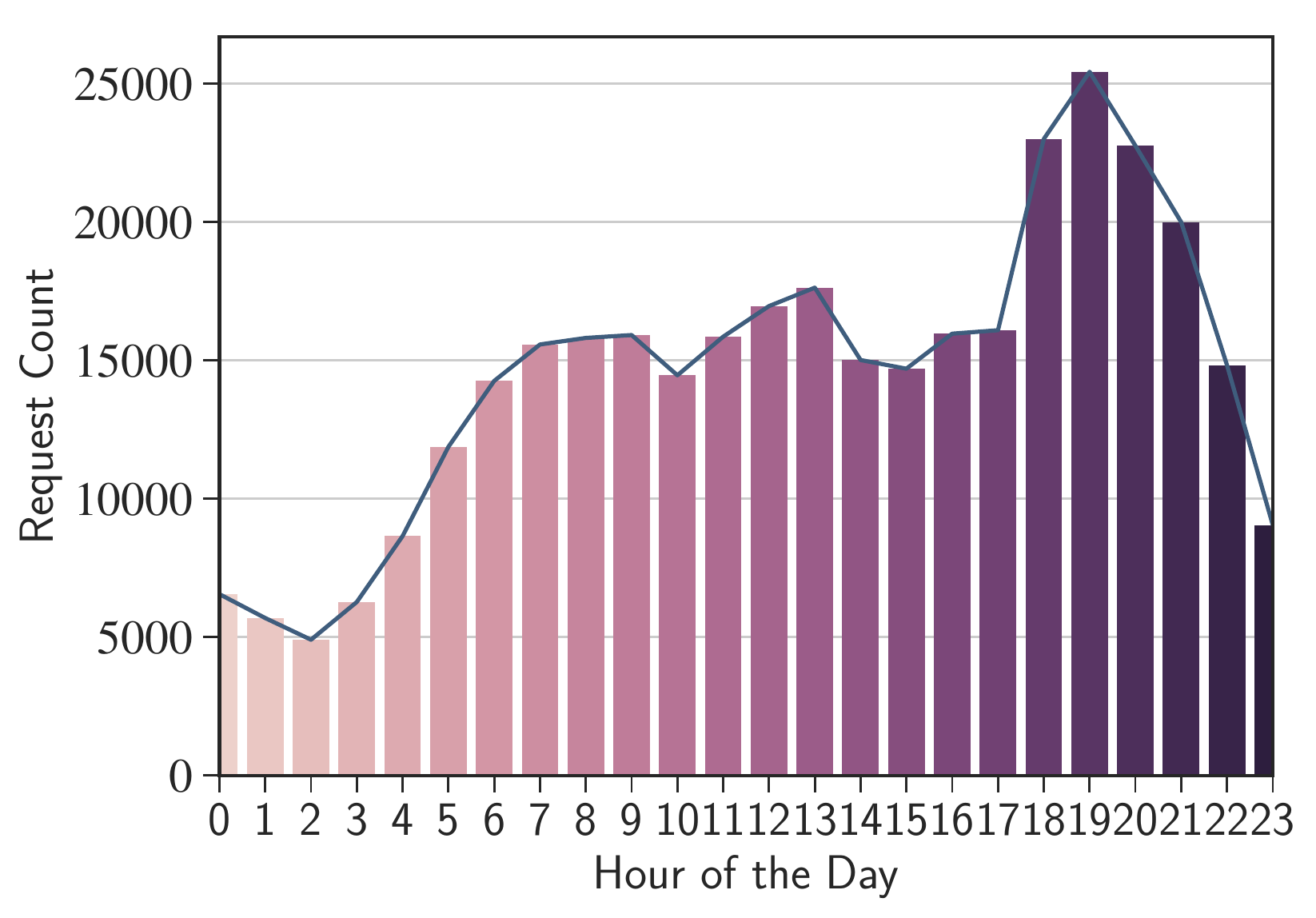}
\caption{Request count}
     \label{fig:hourreqcount}
\end{subfigure}

 \end{tabular}
 \caption{The distribution of request (a),  the request count per square mile of each neighborhood (b), the average traffic speed per hour (c), and the request count per hour (d) in Manhattan on Wednesday, June 8th, 2011.}
\label{fig:density}
\end{figure}

We varied the percentage of user participation from $0\%$ to $100\%$ with an increasing step of $10\%$ to see how different levels of user participation affects the performance of different systems. We used the stratified sampling method~\shortcite{neyman1934two} to select the participating requests because previous research indicates that differences in both neighborhoods and hour of the day can affect passengers' choices of commute modes~\shortcite{jackson1982empirical,schwanen2005affects}. The method first categorized the request data into 888 subgroups according to both spatial differences (37 neighborhoods in Manhattan, see Figure~\ref{fig:neighborhooddensity}) and temporal variations (24 hours, see Figure~\ref{fig:hourreqcount}). For each subgroup, it then randomly selected corresponding percent of requests as the participating requests ($m_{r} =1$) and the rest as the non-participating requests ($m_{r} = 0$).  This produced 11 datasets in total, with each contained the request data for a week.
We considered autonomous ridesharing systems with different settings, including vehicle fleet sizes (1000, 2000, 3000, and 4000), vehicle capacities (1, 2, 4, 6, 8, and 10) and maximum waiting time (2, 4, 6, and 8 minutes). The maximum delay time allowed, including both waiting time and in-car delay time, was calculated by doubling the maximum waiting time (i.e., 4, 8, 12, and 16 minutes).
This resulted in 96 different ridesharing systems. For each system, we performed a simulation on the STARS platform with each of the 11 generated datasets using the anytime optimal algorithm described in~\shortcite{alonso2017demand}. There were 1,056 simulations in total.  For each simulation, we recorded the following information: service rate, average in-car delay ($\delta - \omega$), average waiting time, mean distance traveled by each vehicle during a day, the percentage of shared rides, and the total cost.  For each ridesharing system, we then computed the price of anarchy by assuming $P_{sue}  \subseteq P= \{p_{l}\}$, where $l \in \{0\%, 10\%, 20\%, 30\%, ..., 100\%\}$ is the percentage of user participation, and $p_{l}$ denotes the profile where there are $l$  percent of users that are willing to take a shared ride. Essentially, a profile is a set of all users' decisions on whether to participate in ridesharing or not.

%
%

%
%

We performed another two groups of experiments to study if  user participation affects the efficiency of the same ridesharing system with different request density and different traffic conditions. In the first group, we varied the request density. The typical travel demand in Manhattan are about 3,000,000 trips per week. We compared three kinds of density: half of the demand (x0.5, by removing the even number of requests), the nominal demand (x1.0), and double of the demand (x2.0, by adding the requests of the subsequent week). In the second group, we used the travel time estimate of three different periods in the day: the estimate at 12:00 (lower speed than the average, see Figure~\ref{fig:hourspeed}),  the estimate at 19:00 (higher speed than the average), and the mean daily travel time estimate (average speed). For each condition in each of the group, we ran a simulation on the same ridesharing system (fleet size = 3000, capacity = 4, maximum waiting time = 6 miniutes).

All the simulations were conducted on the same 24-core 3.0GHz Linux machine with 128GB RAM. We used the Gurobi solver (version 7.0.2 with free Academic license) for mixed integer programming optimization. To reduce the total time for simulations,  all the time steps for all simulations were multiplied by a discounting factor $0.1$. The 1,122 simulations took five weeks.

\section{\uppercase{Results}}
This section describes the experimental results, followed by a discussion of their implications. 
\subsection{Main Observations}
Based on the data points from the numerical simulations, we highlight six main observations. We will next discuss each in turn.
\begin{obsv}
User participation typically improves the performance of autonomous ridesharing systems,  but the degree of improvement generally slows down as user participation increases to a high level. 
\end{obsv}
This was observed in all the ARS in terms of total cost, service rate, mean waiting time, percentage of shared rides, and average distance traveled by each vehicle during a day. Due to space limitations, we will report the findings mainly based on the total cost.

\begin{figure*}
\centering
\includegraphics[width=.95\linewidth]{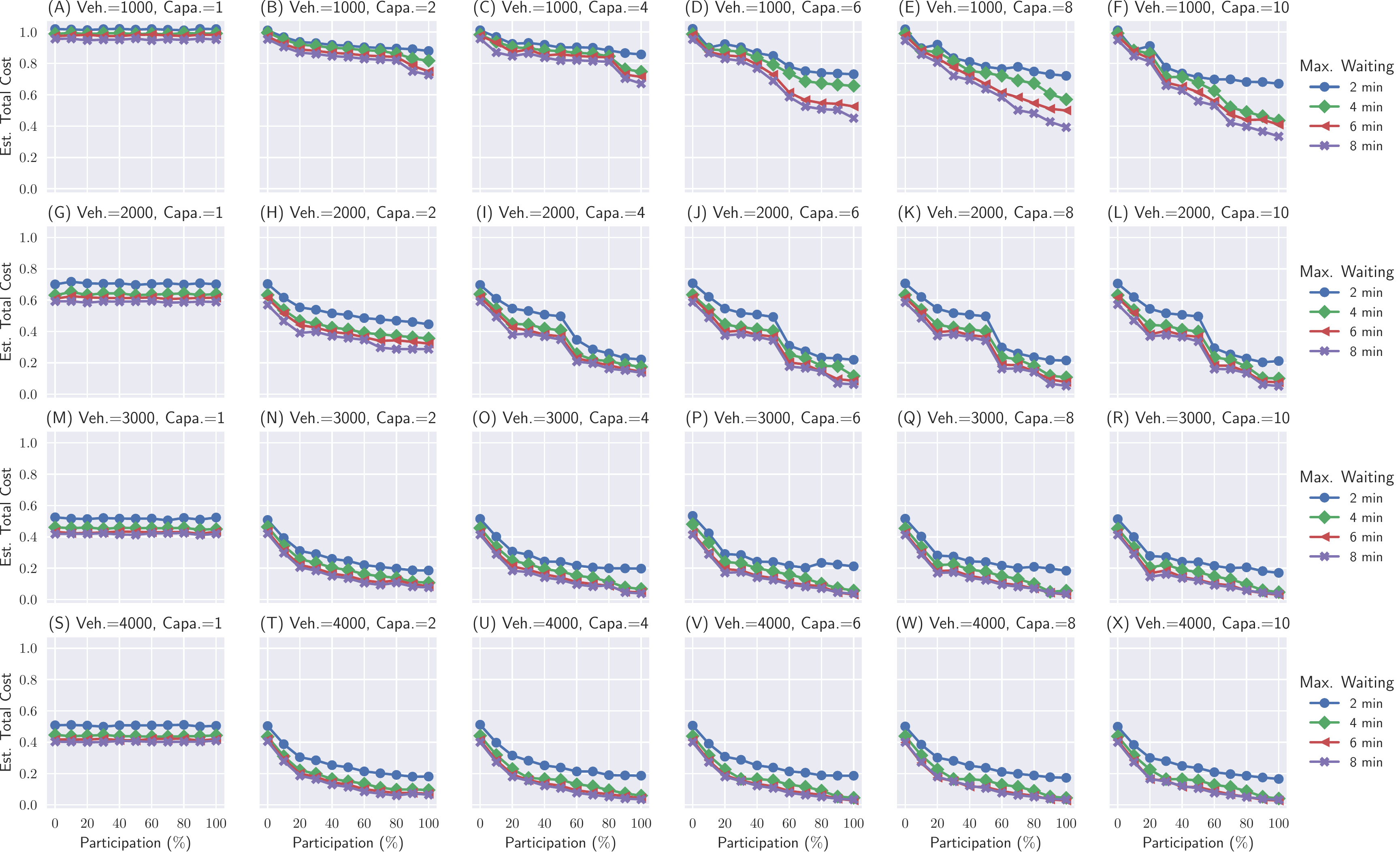}
 \caption{Estimated total cost for ridesharing systems with different configurations.}
      \label{fig:totalcost}
  \end{figure*}

Figure~\ref{fig:totalcost} shows that when the levels of user participation increased, the total cost of all the ridesharing systems (capacity $\geq 2$) decreased. This was expected since a higher degree of user participation provided the rideshairng systems with more room for system-wide optimization: more requests were paired for shared rides and subsequently were serviced. However, when the level of user participation increased to a high level (i.e., above $80\%$), the degree of cost reduction slowed down. This was also expected because when the level of user participation reached to certain level,  substantial improvement by system optimization no longer sustained due to physical limits of the transportation systems such as spatial and temporal factors~\cite{tsao1999spatial}.

Note that systems in which the vehicles have the capacity of 1 are excluded since they are not capable of offering carpooling services. In our work, we listed the performance of these systems for comparison purpose only. 

%
%
%
%

\begin{obsv}
User participation typically has a greater impact on the performance of autonomous ridesharing systems with small fleets of high-capacity vehicles.
\end{obsv}
Figure~\ref{fig:totalcost} shows that the ARS experienced a larger degree of cost reduction when they were equipped with small fleets of high-capacity vehicles. For instance, higher degree of performance improvement was observed in the ridesharing systems with the following conditions:  (1) Vehicle Size = 1000, Capacity = 8 (See Figure~\ref{fig:totalcost}E); (2) Vehicle Size = 1000, Capacity = 10 (See Figure~\ref{fig:totalcost}F); (3) Vehicle Size = 2000, Capacity = 8 (See Figure~\ref{fig:totalcost}K); and (4) Vehicle Size = 2000, Capacity = 10 (See Figure~\ref{fig:totalcost}L). This trend implies that participation tends to have a greater impact on the performance of ARS with small fleets of high-capacity vehicles than others. An explanation for this trend is that the service rates for the former systems were initially quite low ($20\%$ --  $40\%$) due to small fleet sizes and no ridesharing participants.  As the level of user participation increases, the large capacity of the vehicles allowed a higher degree of ride sharing, making the degree of performance improvement substantially higher than others.

     \begin{figure}
\centering
\includegraphics[width=.65\linewidth]{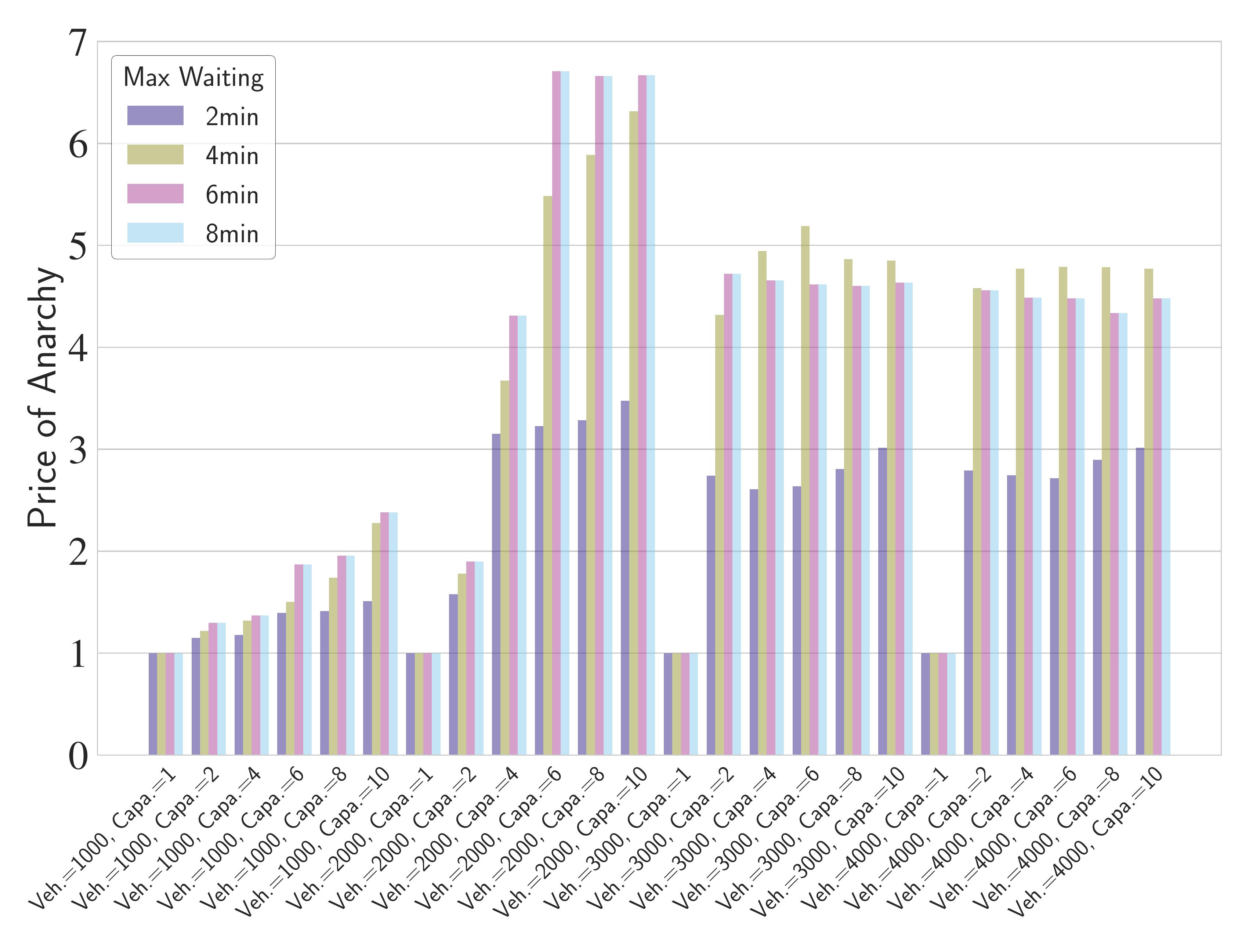}
\caption{Price of anarchy for ridesharing systems with different configurations.}
     \label{fig:poa}
\end{figure}

\begin{obsv}
User participation typically has a smaller impact on the performance of autonomous ridesharing systems with a short period of maximum waiting time allowed.
\end{obsv}  
This was also observed in ARS with all the configurations. For example, Figure~\ref{fig:totalcost}B demonstrates that systems with maximum waiting time of 2 minutes had a lower degree of cost cutting than the other three systems. The cost of the system with maximum waiting time of 2 minutes dropped from 1.0 to 0.88 as the participation level increased from $0\%$ to $100\%$,  while the system with maximum waiting time of 8 minutes experienced a greater reduction (from 1.0 to 0.72) in the total cost.  This reason is that a looser constraint on maximum waiting time enabled the systems to service more requests,   which was not possible if given a narrow time frame. Similar patterns were also found when the vehicle capacity varied while other settings kept unchanged:  systems with larger capacity tended to have a larger shrinkage in the cost.  

\begin{obsv}
As the fleet size increases, the price of anarchy due to users' uncoordinated choices on participation typically first increases and then decreases.
\end{obsv} 

Figure~\ref{fig:poa} shows that as the fleet size increased, the price of anarchy first rose steeply (from Vehicle Size = 1000 to 2000) and then dropped to a level with small fluctuations (from Vehicle Size = 2000 to 4000).  Specifically, when the fleet size was small ($1000$), the price of anarchy climbed gradually as the vehicle capacity increased.  When the fleet size increased to 2000, the price of anarchy jumped more than doubled when the vehicle capacity was switched from 2 to 4, for all conditions of maximum waiting time. Several factors contributed to this sharp rise: the service rate of the system with vehicle capacity 2  experienced a larger boost than that of the system with vehicle capacity 4, while the changes in the average in-car delay and the mean waiting time kept the same pace.

It is interesting to note that when the fleet size was large ($\geq 3000$), the price of anarchy fluctuated slightly as the vehicle capacity changed.  The price of anarchy for systems with a fleet size of 3000 was almost the same as the systems with a fleet of 4000 vehicles. The reason is that the service rates for both conditions were  initially at the same level (about $60\%$),  and systems with both configurations were sufficient for satisfying the travel demand of daily commuters in Manhattan. This implies that a fleet of 3000 vehicles might serve the travel demand of passengers in Manhattan reasonably well. This was consistent with the results by~\shortciteN{alonso2017demand}. It further indicates that appropriate configurations  of the autonomous  ridesharing systems can be set to keep the inefficiency caused by passengers' uncoordinated behavior on ridesharing participation minimal.


\begin{figure*}[h]
\centering
\begin{tabular}{cccccc}
\begin{subfigure}[b]{.15\linewidth}
  \includegraphics[width=\linewidth]{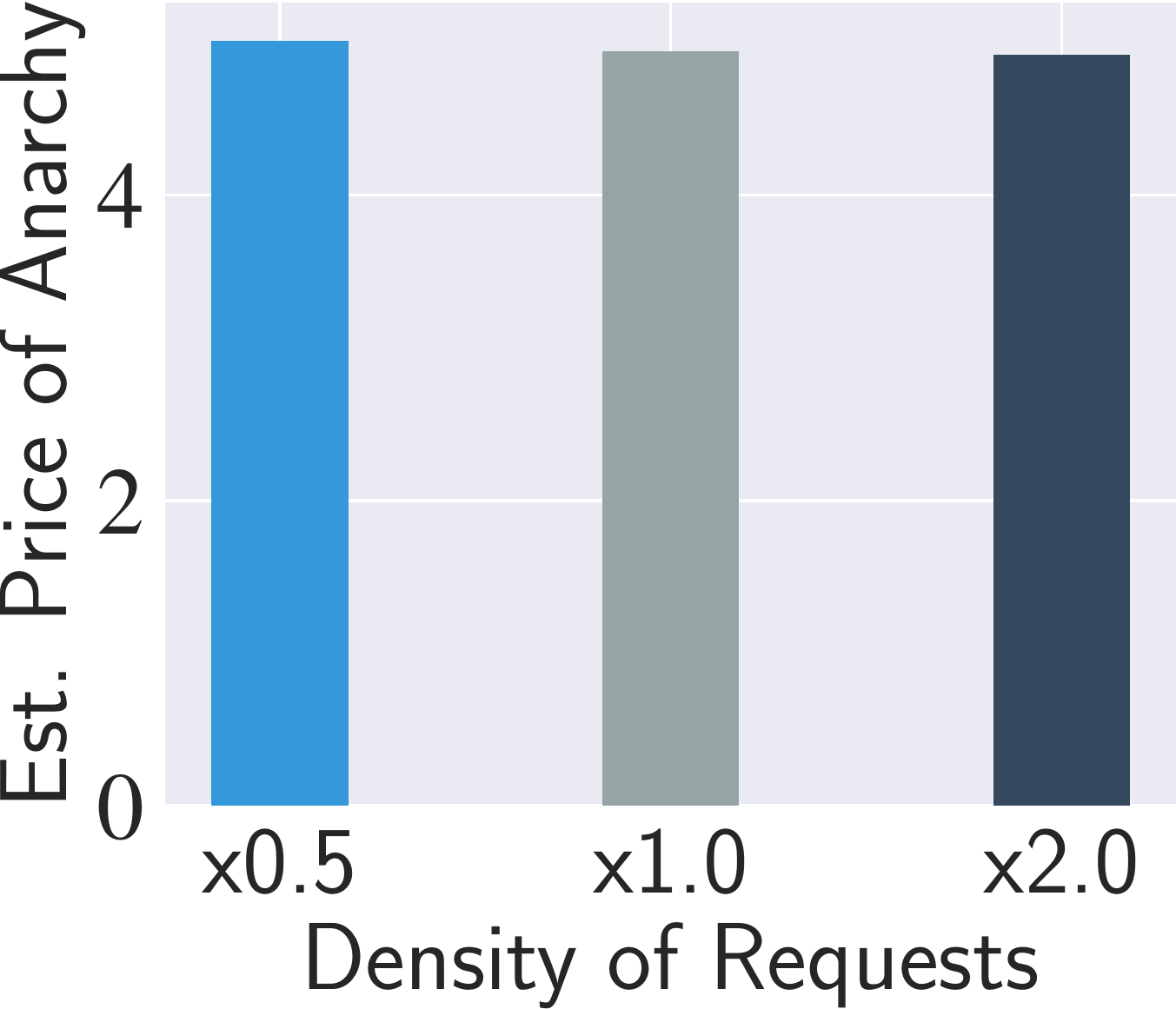}
    \caption{PoA}
      \label{fig:densitypoa}
\end{subfigure}

\begin{subfigure}[b]{.16\linewidth}
  \includegraphics[width=\linewidth]{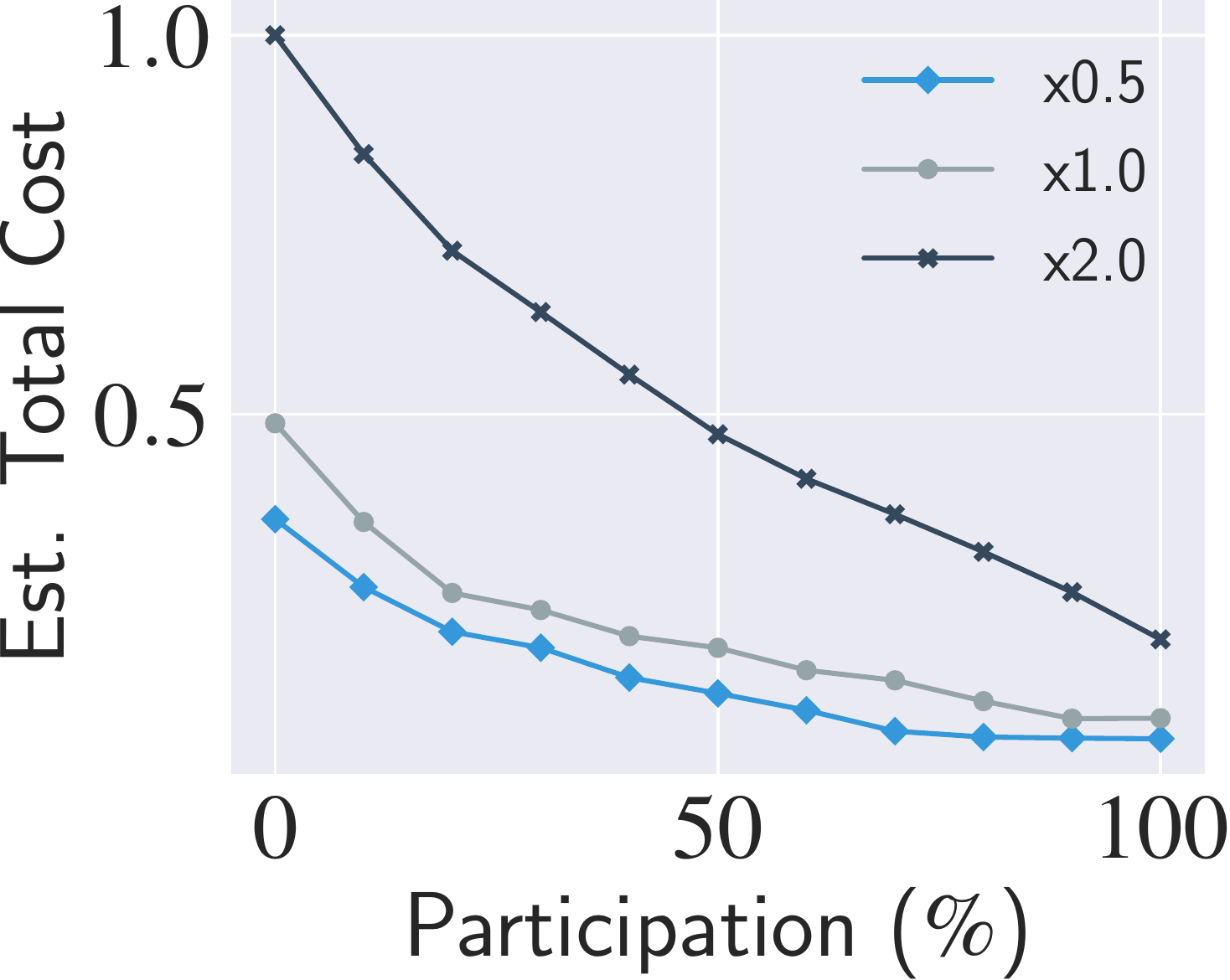}
    \caption{Est. cost}
      \label{fig:densitycost}
\end{subfigure}

\begin{subfigure}[b]{.16\linewidth}
  \includegraphics[width=\linewidth]{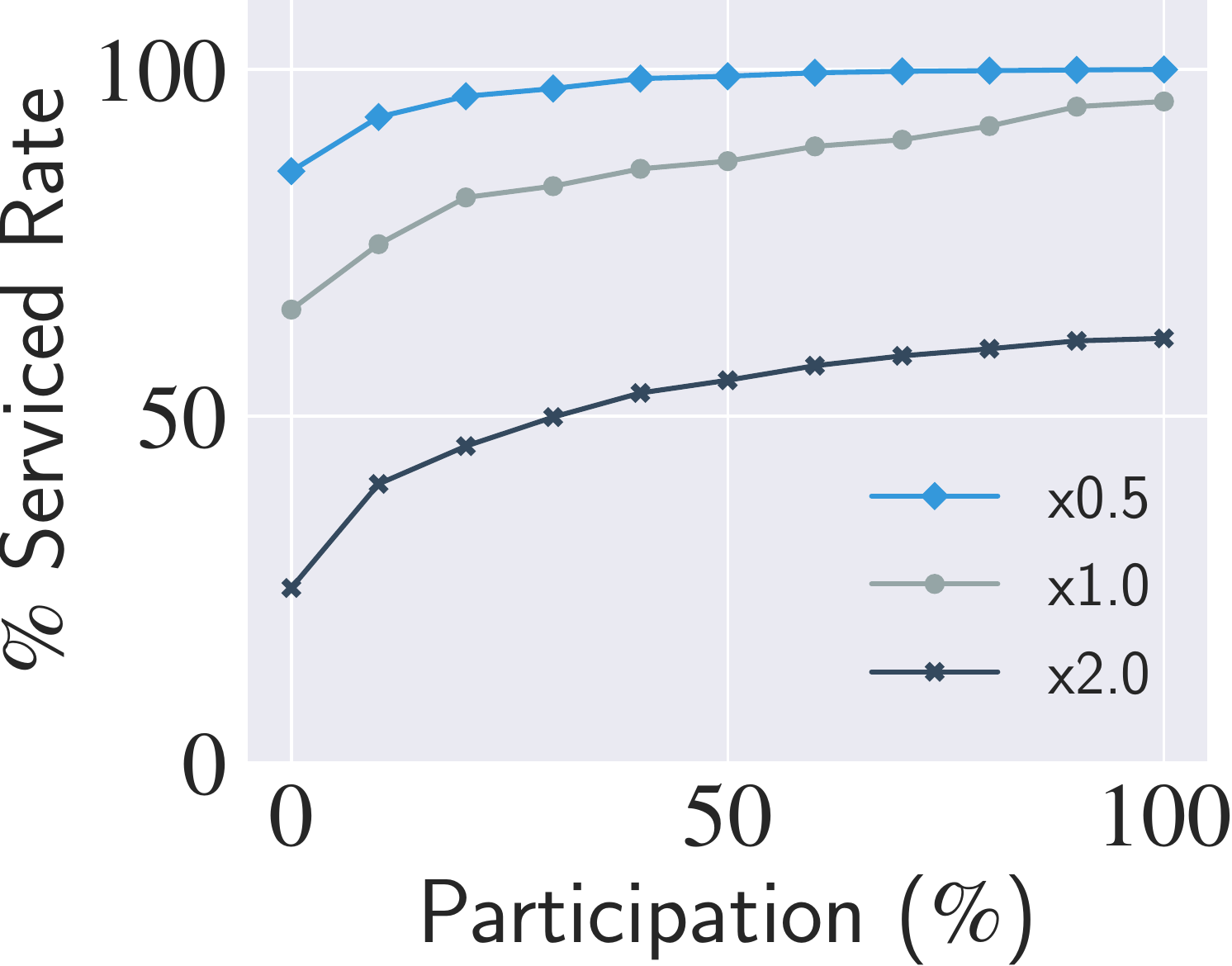}
    \caption{Service rate}
      \label{fig:densityservice}
\end{subfigure}

\begin{subfigure}[b]{.16\linewidth}
  \includegraphics[width=\linewidth]{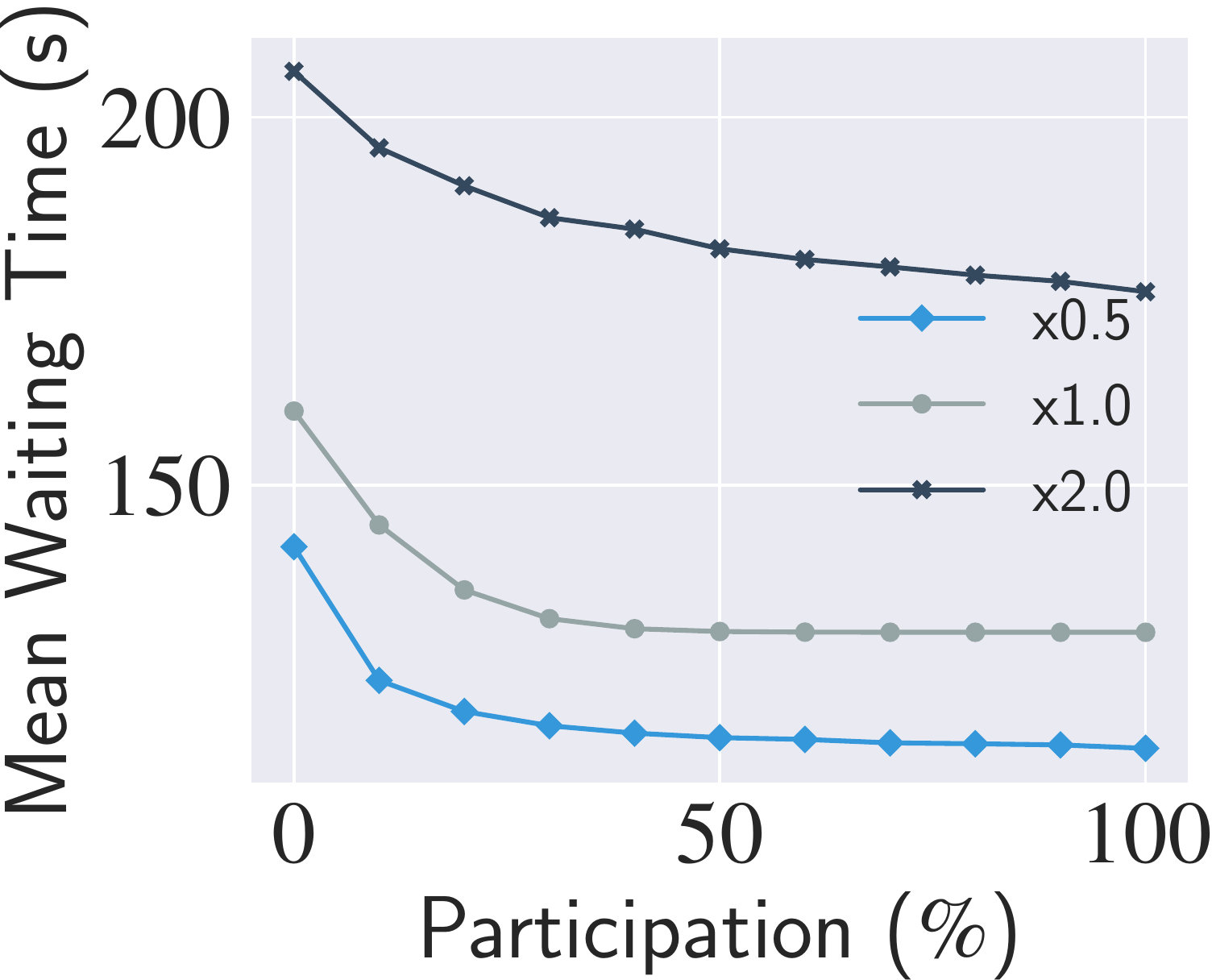}
    \caption{Mean waiting}
      \label{fig:densitywaiting}
\end{subfigure}

\begin{subfigure}[b]{.16\linewidth}
  \includegraphics[width=\linewidth]{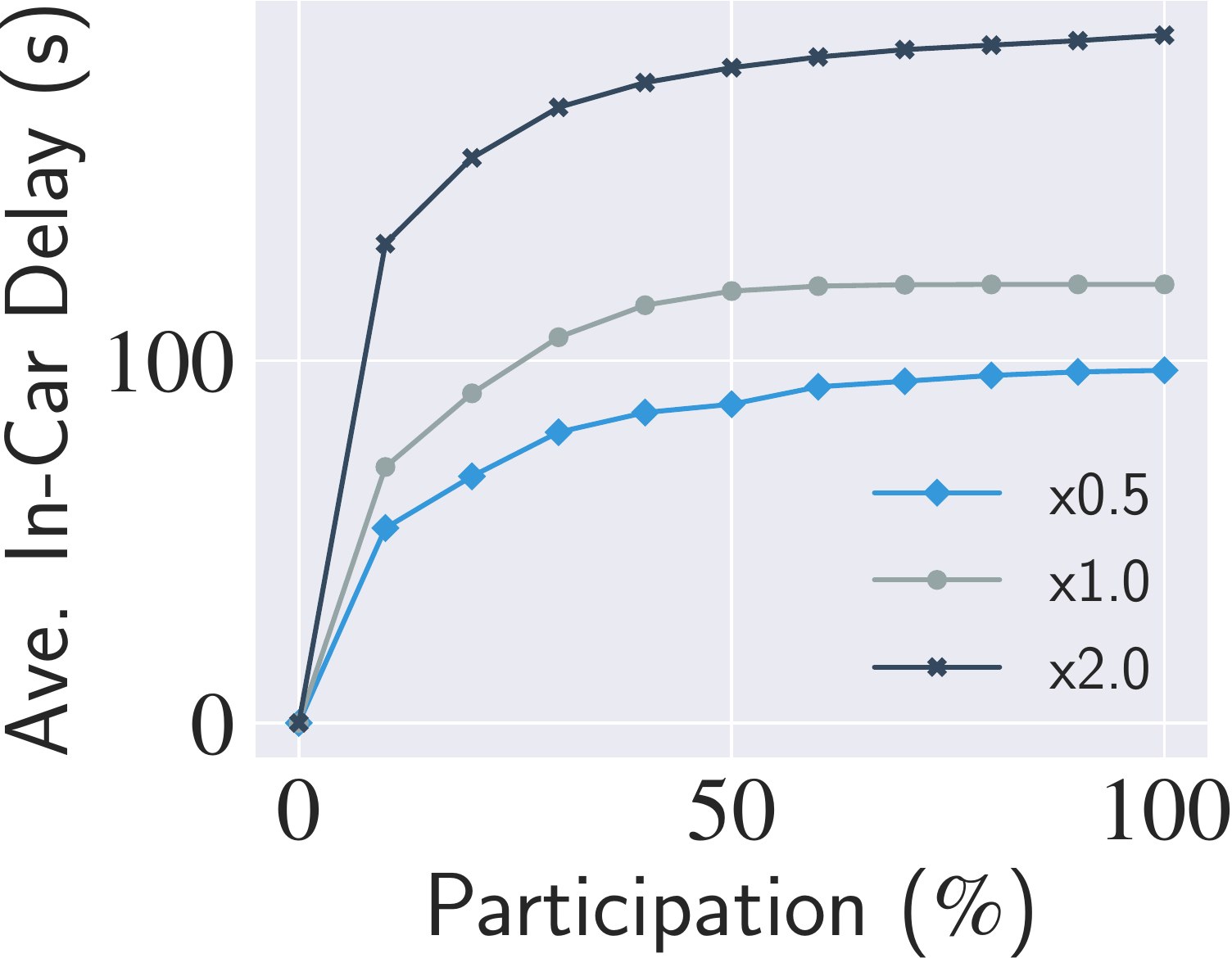}
    \caption{Mean delay}
      \label{fig:denpoa}
\end{subfigure}

\begin{subfigure}[b]{.15\linewidth}
  \includegraphics[width=\linewidth]{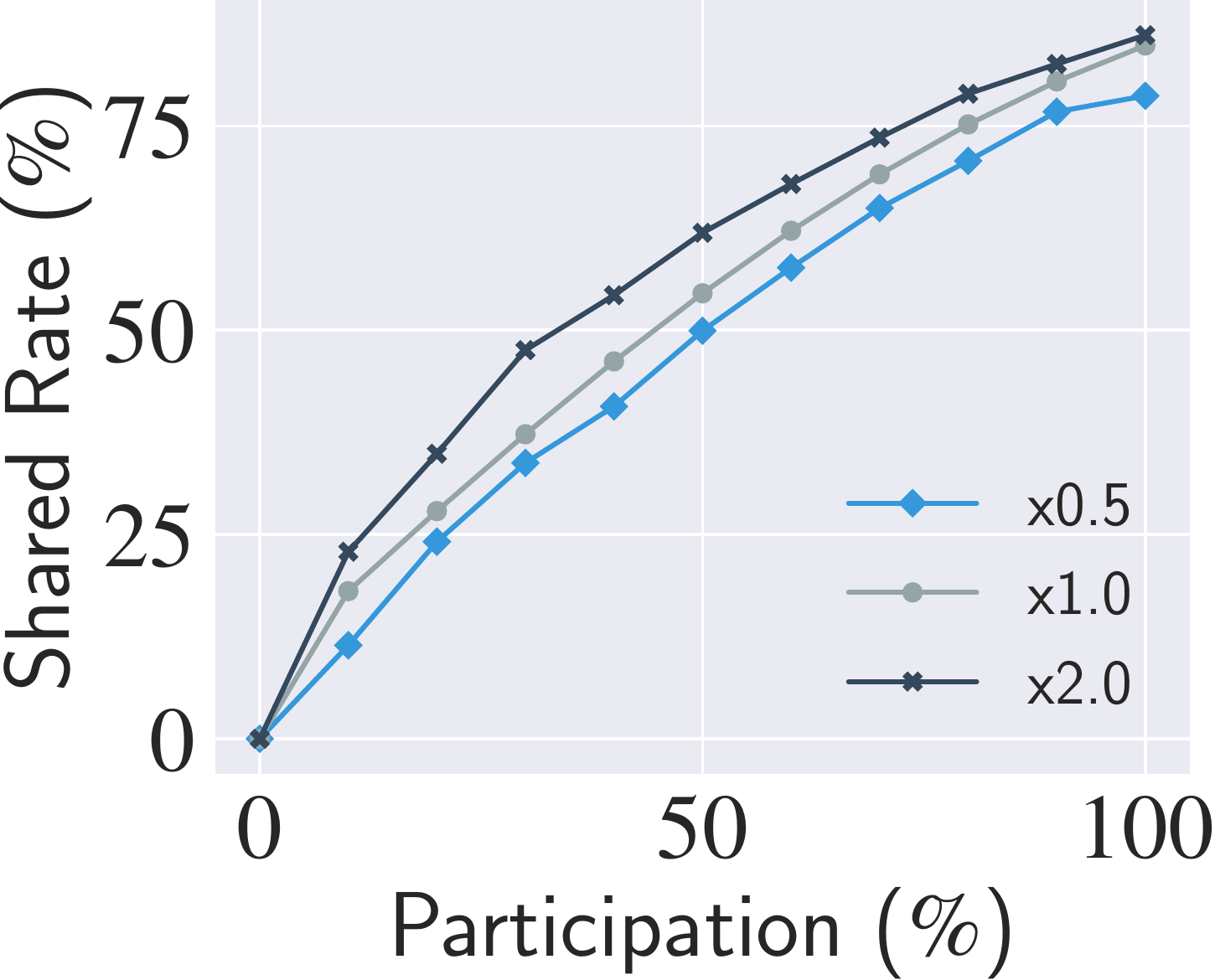}
    \caption{Shared rate}
      \label{fig:densityshare}
\end{subfigure}
\end{tabular}
\caption{A comparison of the price of anarchy and efficiency on a ridesharing system  (fleet size =3000, capacity=4, and maximum waiting time = 6 minutes) by varying request density.}
\label{fig:varydensity}
\end{figure*}
\begin{figure*}[h]
\centering
\begin{tabular}{cccccc}
\begin{subfigure}[b]{.15\linewidth}
  \includegraphics[width=\linewidth]{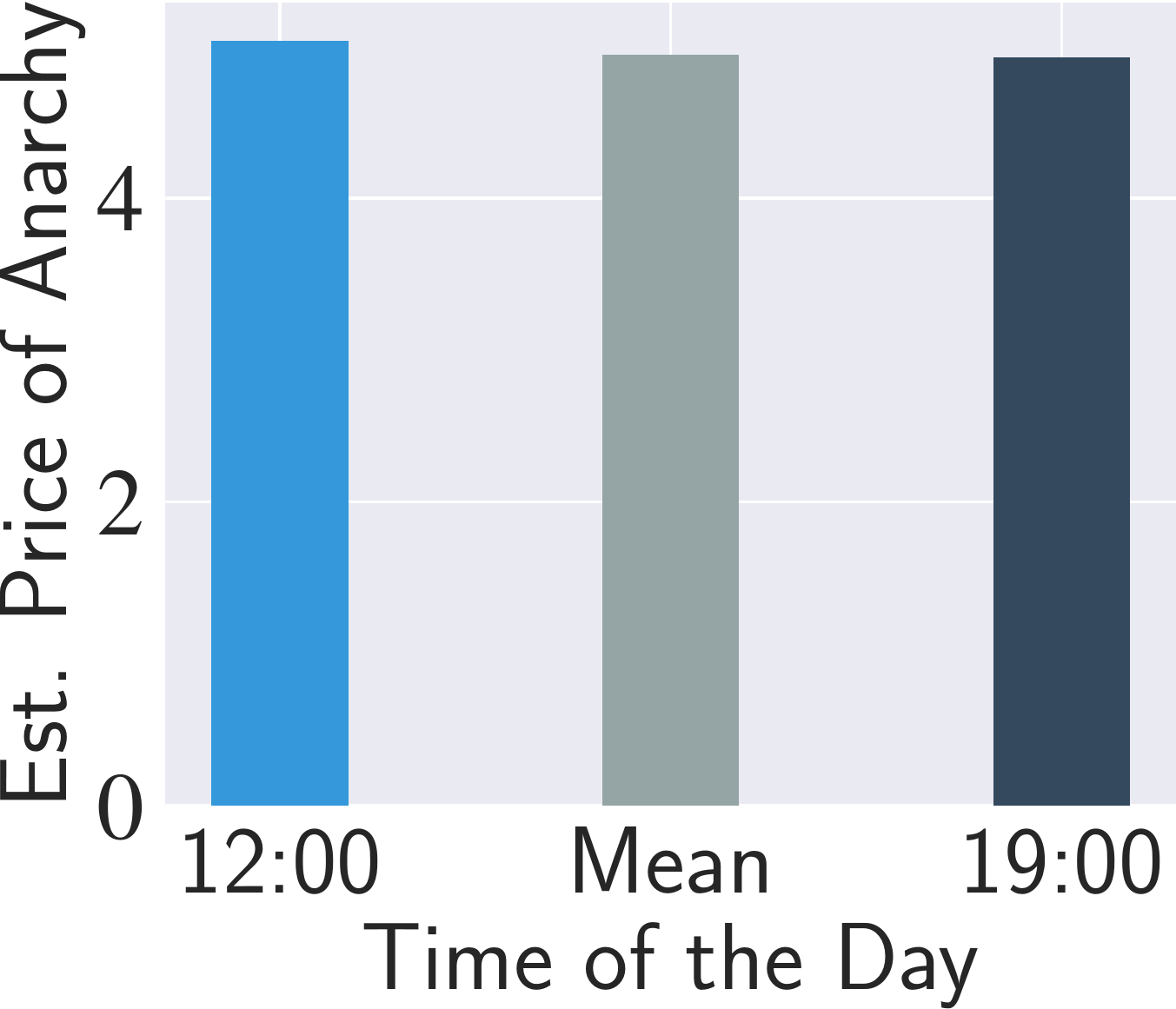}
    \caption{PoA}
      \label{fig:congestionpoa}
\end{subfigure}

\begin{subfigure}[b]{.16\linewidth}
  \includegraphics[width=\linewidth]{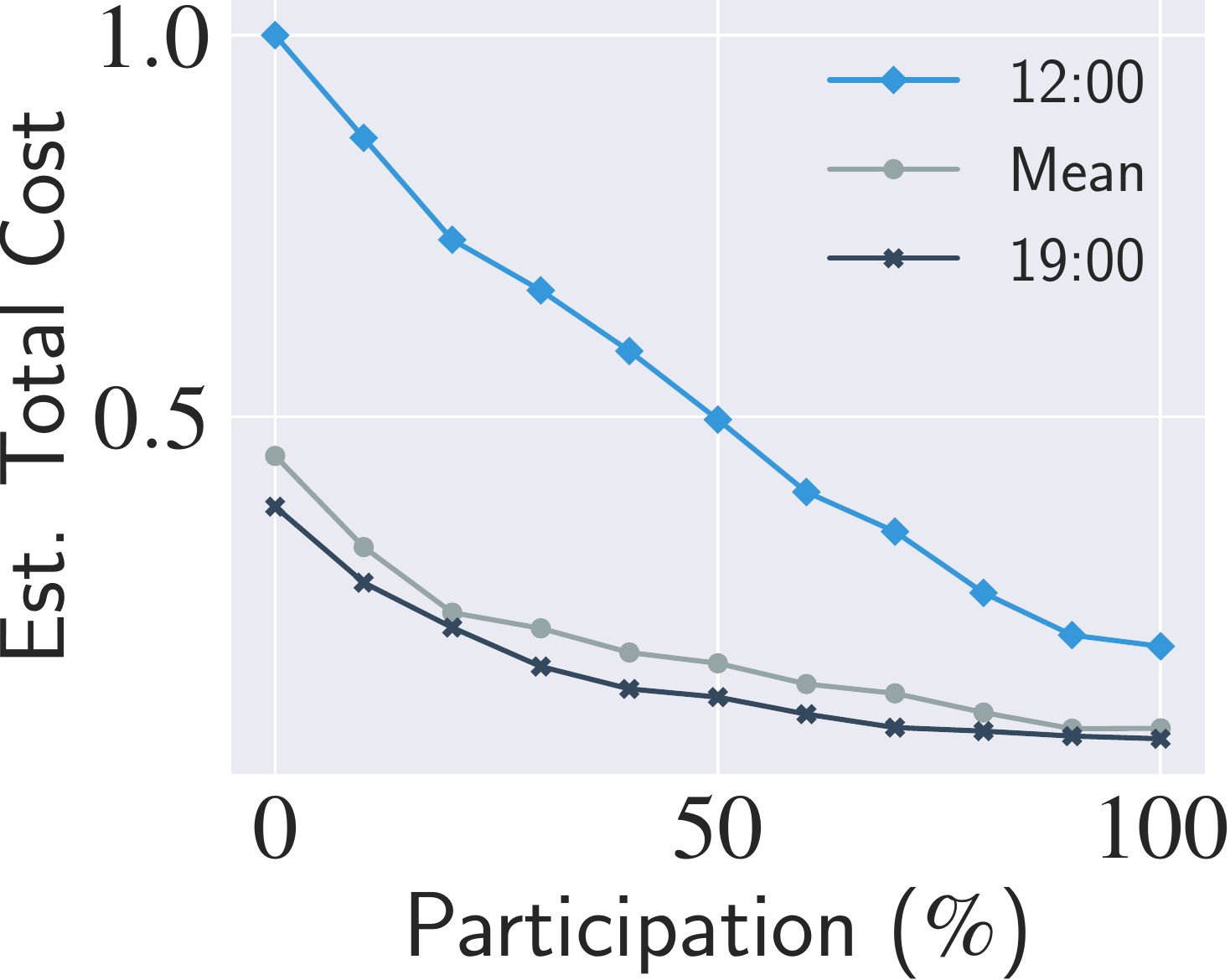}
    \caption{Est. cost}
      \label{fig:congestioncost}
\end{subfigure}

\begin{subfigure}[b]{.16\linewidth}
  \includegraphics[width=\linewidth]{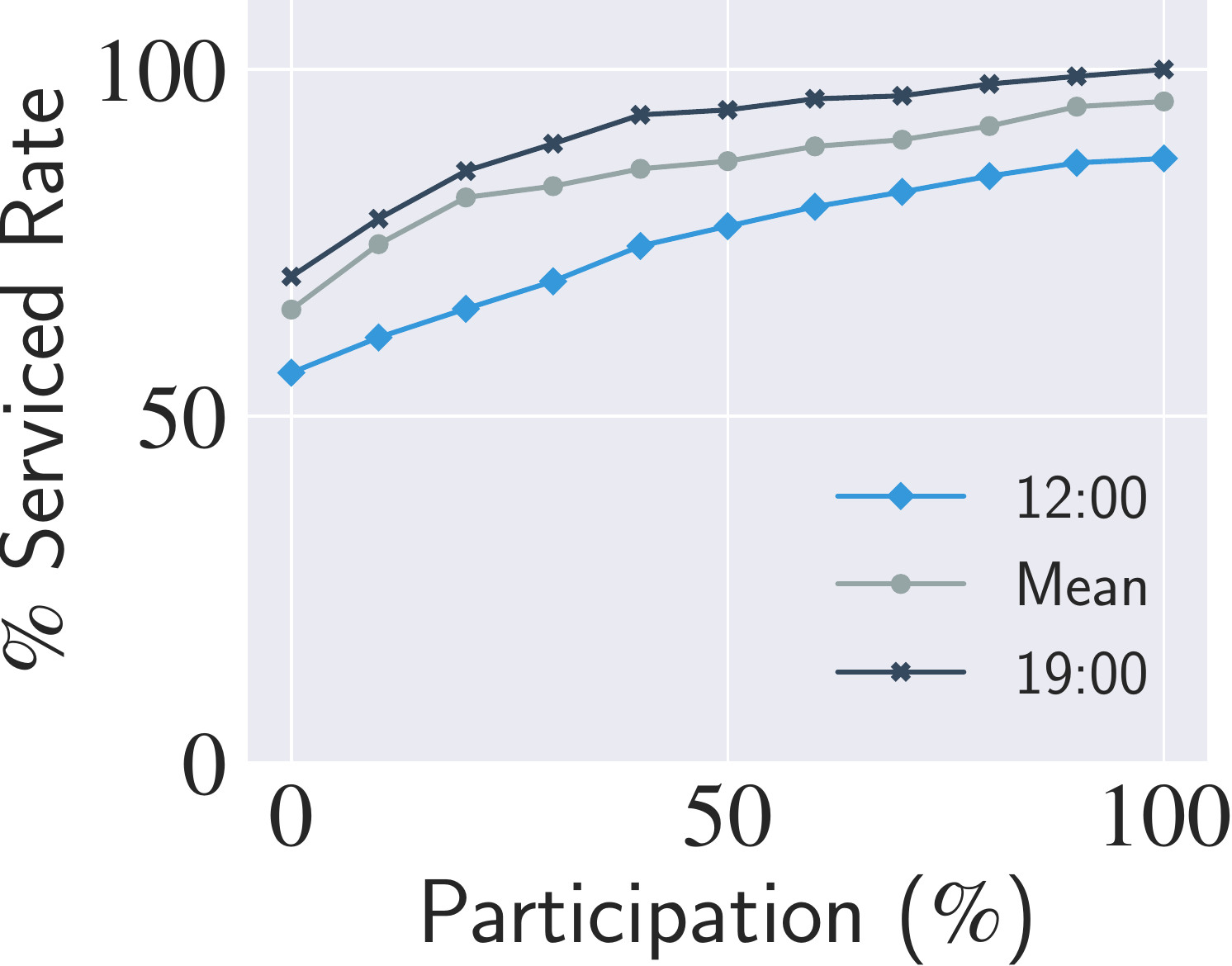}
    \caption{Service rate}
      \label{fig:congestionservicerate}
\end{subfigure}

\begin{subfigure}[b]{.16\linewidth}
  \includegraphics[width=\linewidth]{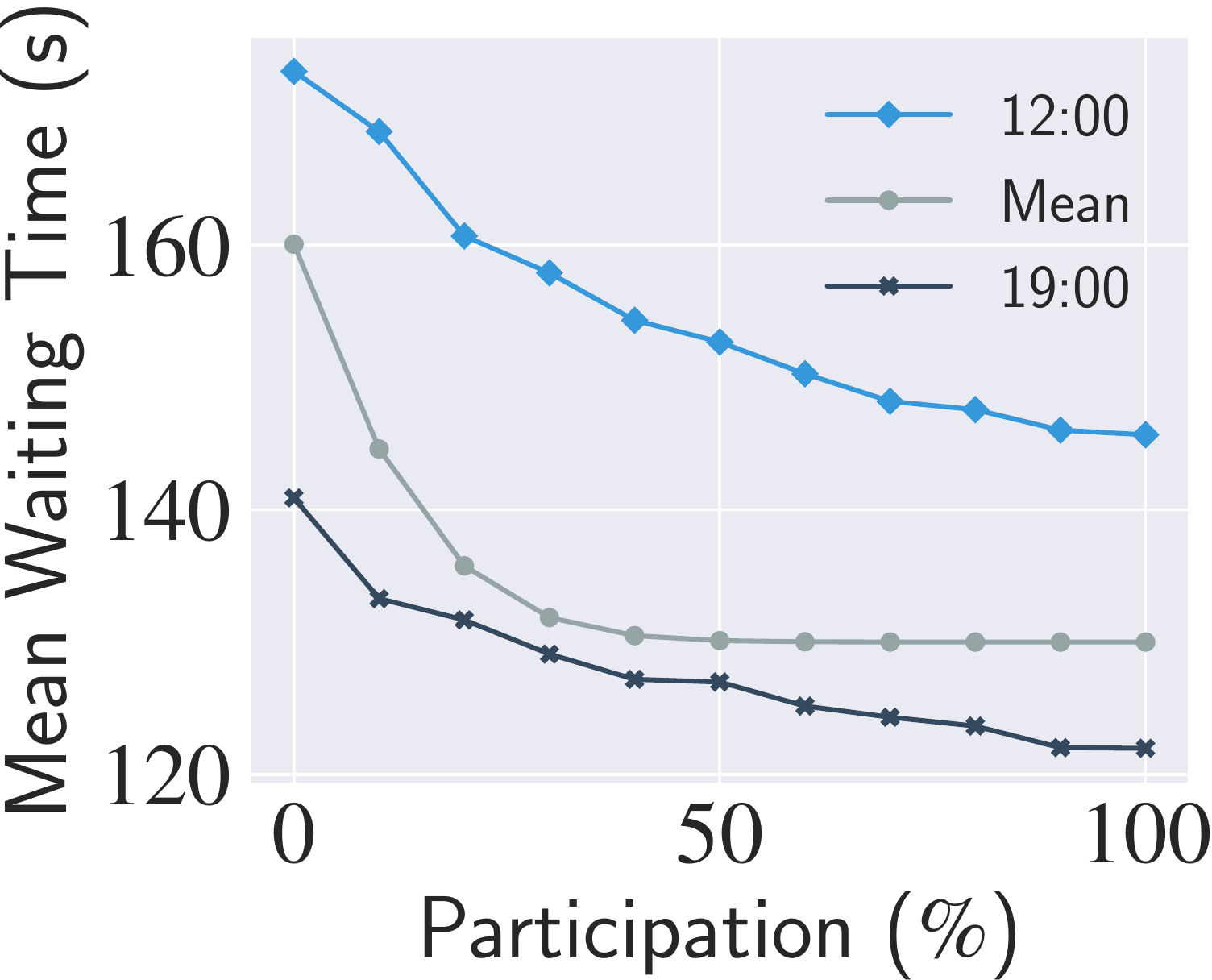}
    \caption{Mean waiting}
      \label{fig:congestionwaiting}
\end{subfigure}

\begin{subfigure}[b]{.16\linewidth}
  \includegraphics[width=\linewidth]{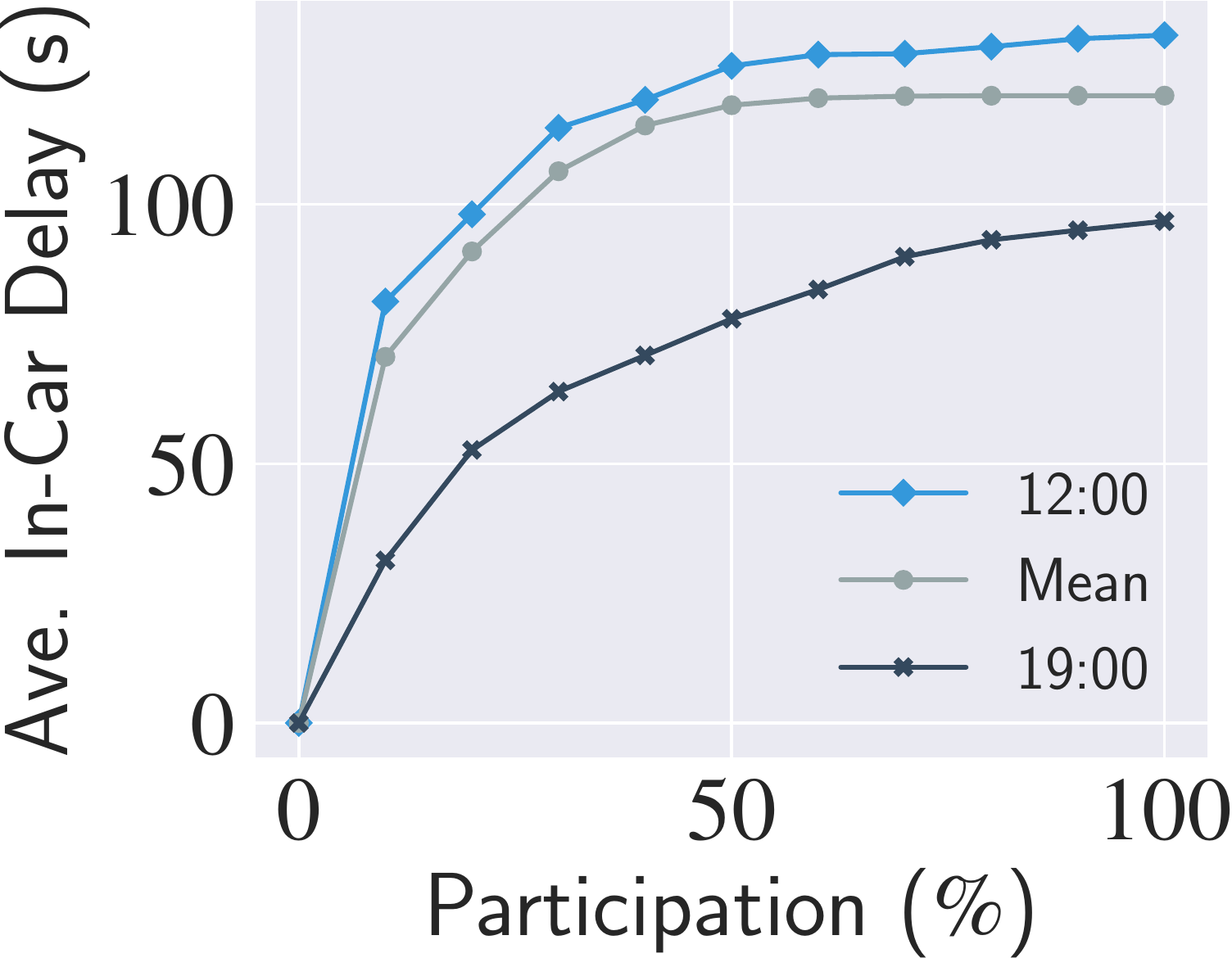}
    \caption{Mean delay}
      \label{fig:congestionincardelay}
\end{subfigure}

\begin{subfigure}[b]{.15\linewidth}
  \includegraphics[width=\linewidth]{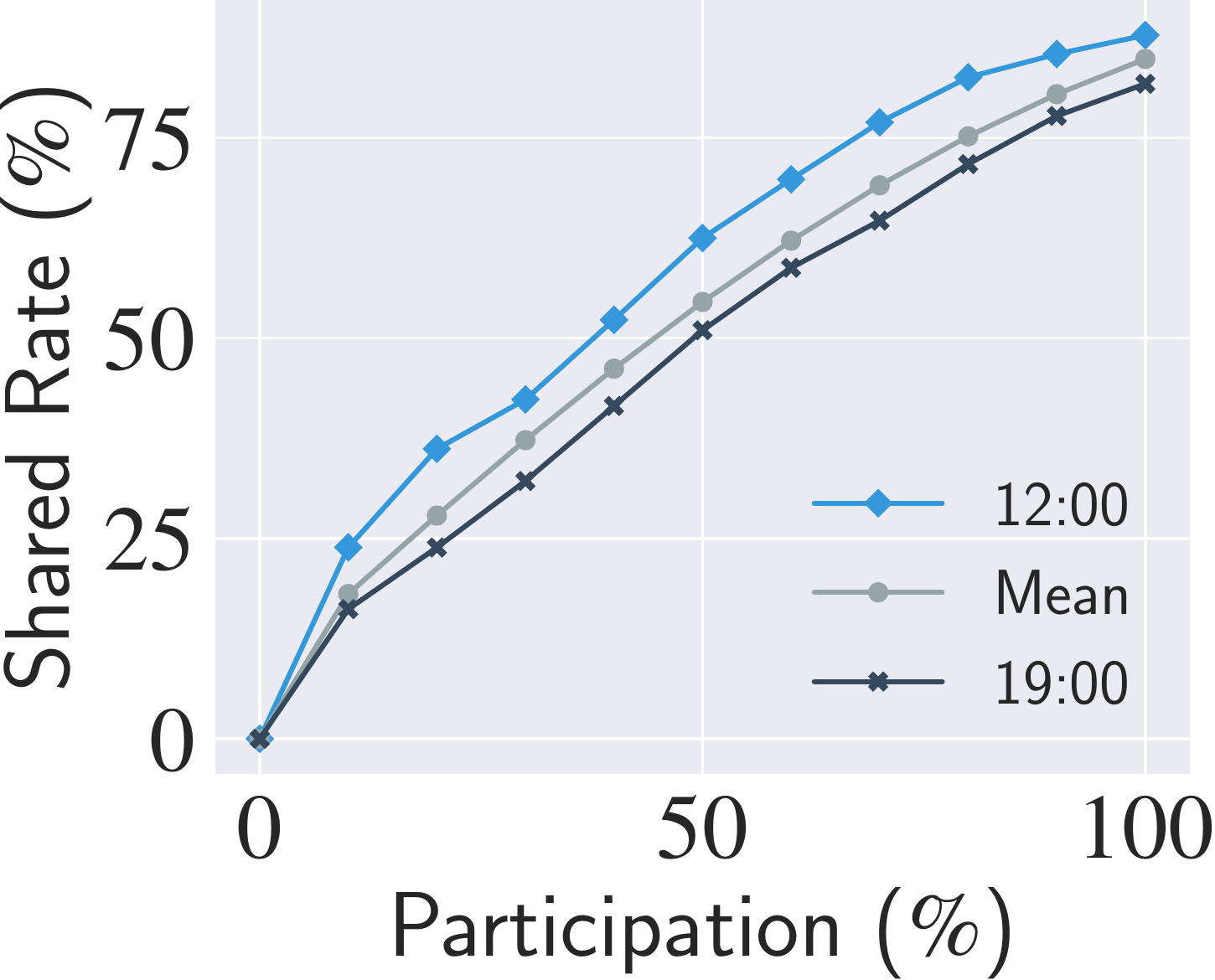}
    \caption{Shared rate}
      \label{fig:congestionshare}
\end{subfigure}
\end{tabular}
\caption{A comparison of the price of anarchy and efficiency on a ridesharing system  (fleet size =3000, capacity=4, and maximum waiting time = 6 minutes) by  varying the time of the day used for travel time estimate.}
\label{fig:varytime}
\end{figure*}

\begin{obsv}
The price of anarchy due to users' uncoordinated choices on participation typically keeps steady when the maximum waiting time is sufficiently long.
\end{obsv}
%

When the fleet size was large ($\geq 3000$), the PoA experienced a significant increase (more than $50\%$) when switching the maximum waiting time from 2 minutes to 4 minutes. This was expected since the service rates of systems with both were almost the same,  while systems with maximum waiting time of 4 minutes had looser constraints and allowed more room the ride matching than the other. As a result, the former experienced a greater improvement of system efficiency. 

Interestingly, as the maximum waiting time increased from 6 minutes to 8 minutes, the price of anarchy stayed almost the same despite variations on vehicle capacity or fleet size. The reason is that when the maximum waiting time reaches a sufficient long duration (e.g., 6 minutes), additional waiting time does not improve the performance of ridesharing systems significantly. Other parameters become critical. This implies that it is possible to reduce the influence of user participation on the performance of ridesharing systems by identifying proper maximum waiting time.




\begin{obsv}
Request density or traffic condition typically has little impact on the price of anarchy, although the system efficiency generally boosts when the traffic condition improves or the request density reduces.
\end{obsv}

Figure~\ref{fig:varydensity} demonstrates that user participation had similar impact on ridesharing systems with different request density. Specifically,  when the level of user participation increased, the total cost decreased (See Figure~\ref{fig:densitycost}). This was expected since higher levels of user participation typically increase the chance of successful ride matching (See Figure~\ref{fig:densityservice}) and shared rate (See Figure~\ref{fig:densityshare}). 

Noticeably,  the price of anarchy remained almost unchanged (approximately 5.0) as the density increased (See Figure~\ref{fig:densitypoa}).  The reason is described as follows: from Figure~\ref{fig:densityservice}, one can see that the service rates of systems with half of the demand (x$0.5$), and nominal demand (x$1.0$) were fairly close, and were higher than the service rate of the system with double demand (x$2.0$). However, they grew slower than the service rate of the system with double demand. Although the latter system performed worse than the former two in terms of mean waiting time and mean in-car delay, the ratios between the total cost of the worst state (participation level = $0\%$) and the best state (participation level = $100\%$) were the same.

 Figure~\ref{fig:varytime} illustrates that as the travel condition improved, the system efficiency increased.  Similarly, the price of anarchy did not vary according to traffic condition.
 
\subsection*{Discussion}
All the six findings tell a similar story: there are sweet spots (e.g., Vehicle Size = 3000, Capacity = 6, Maximum Waiting Time = 6 minutes) that can be utilized to make tradeoffs to keep the price of anarchy minimal while maintaining a good system efficiency. Stakeholders of ridesharing systems should base decisions regarding system configurations on insights gained from realistic simulations with real-world data when experiments on deployed systems are not possible or too costly.

Our work distinguishes itself from game-theoretic approaches in the sense that we do not assume passengers share the same utility function as many works~\shortcite{roughgarden2005selfish,christodoulou2005price,shen2016online} do. In particular, we do not make the assumption that passengers' utilities on participation are determined by a unified function or drawn from a probability distribution. This is because passengers' choices on commuting modes are usually influenced by many factors, such as time variability~\shortcite{jackson1982empirical}, neighborhoods~\shortcite{schwanen2005affects},  habits and prior experiences~\shortcite{gargiulo2015dynamic}. It is rather difficult or even unrealistic to isolate a utility function or a probability distribution that characterizes all passengers' decision-making processes. 

While illuminating, we must be careful not to overgeneralize these results. Although we conducted 1,122 simulations on ridesharing systems with  96 configurations, our work is an exploratory study rather than an exhaustive one. Similarly, when sampling the ride requests as the ridesharing participants, it was not possible for us to include all the combinations.  Our work is intended to spur a pipeline of further studies to deepen our understanding of how human dynamics impact the efficiency of large social-technological systems, as well as to shed light on integrating advances in programming languages into the development of scalable, data-driven,  multi-agent based applications. Future work is needed to identify how to best do so.

\section{\uppercase{Conclusion}}
In this paper, we describe a modular design for agent-based simulations of ridesharing systems. We introduce a multi-agent simulation platform called STARS, for simulating city-scale, autonomous ridesharing systems. Via extensive experiments on STARS with real-world datasets in ridesharing, we present the first simulation analysis toward quantifying the impact of users' uncoordinated behavior on participation in autonomous ridesharing systems.  The two messages to take away are: (1) specific configurations (e.g., fleet size, vehicle capacity, and the maximum waiting time) of ridesharing systems can be identified to counter the effect of user participation on the system efficiency; and (2) tradeoffs between system efficiency and price of anarchy are needed and are often feasible  to achieve desired outcomes.

There are a number of interesting avenues for further research. One of them is to investigate whether our results will generalize to autonomous ridesharing systems with different road networks. Another one is to develop novel approaches (e.g., incentive mechanisms, information structures and regulations) to reduce the price of anarchy in autonomous ridesharing systems.  We also find it very rewarding to explore new methods for scalable multi-agent simulations.

\section*{ACKNOWLEDGMENTS}
This work was supported by National Science Foundation through grant no. CCF-1526593.

\bibliographystyle{wsc}
\bibliography{amod}

\appendix

\section*{APPENDIX} 
\label{sec:append}

\begin{figure*}
  \centering
  \includegraphics[width=\linewidth]{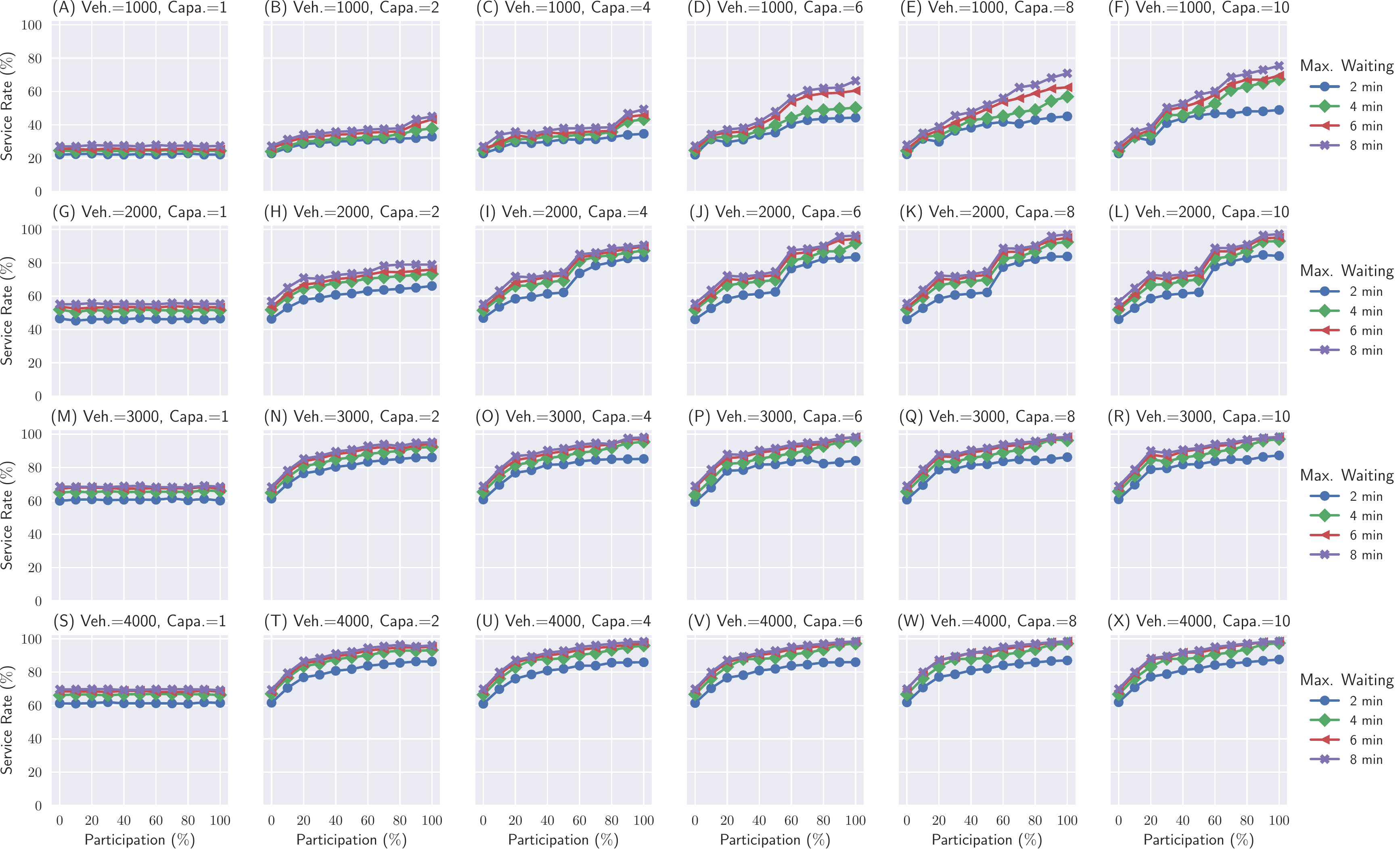}
    \caption{Service rate (percentage of the requests serviced) for autonomous ridesharing systems with different configurations.}
     \label{fig:servicerate}

     \end{figure*}
\begin{figure*}
  \centering
  \includegraphics[width=\linewidth]{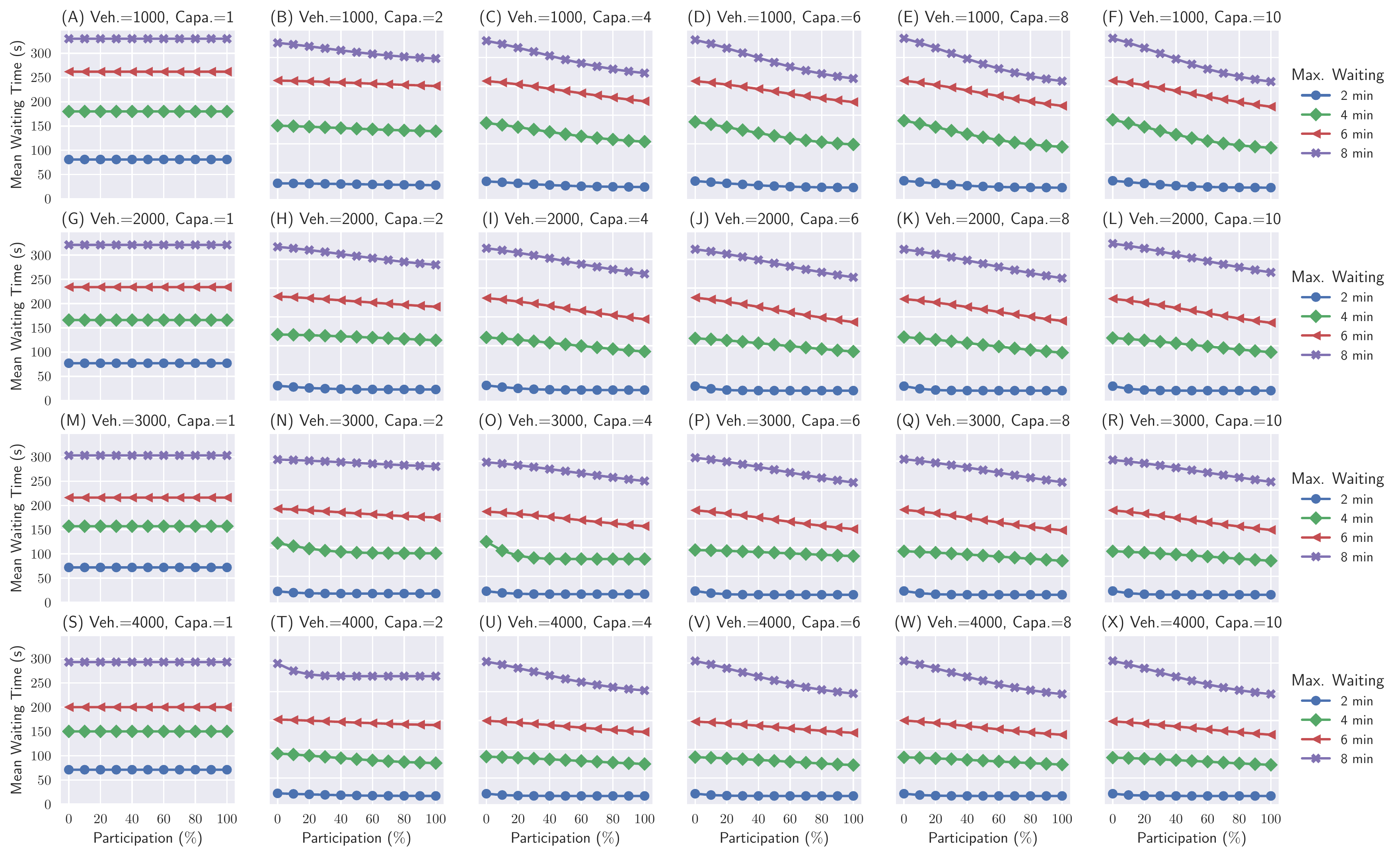}
    \caption{Mean waiting time for autonomous ridesharing systems with different configurations.}
      \label{fig:meanwaiting}
\end{figure*}

\begin{figure*}
\centering
\includegraphics[width=\linewidth]{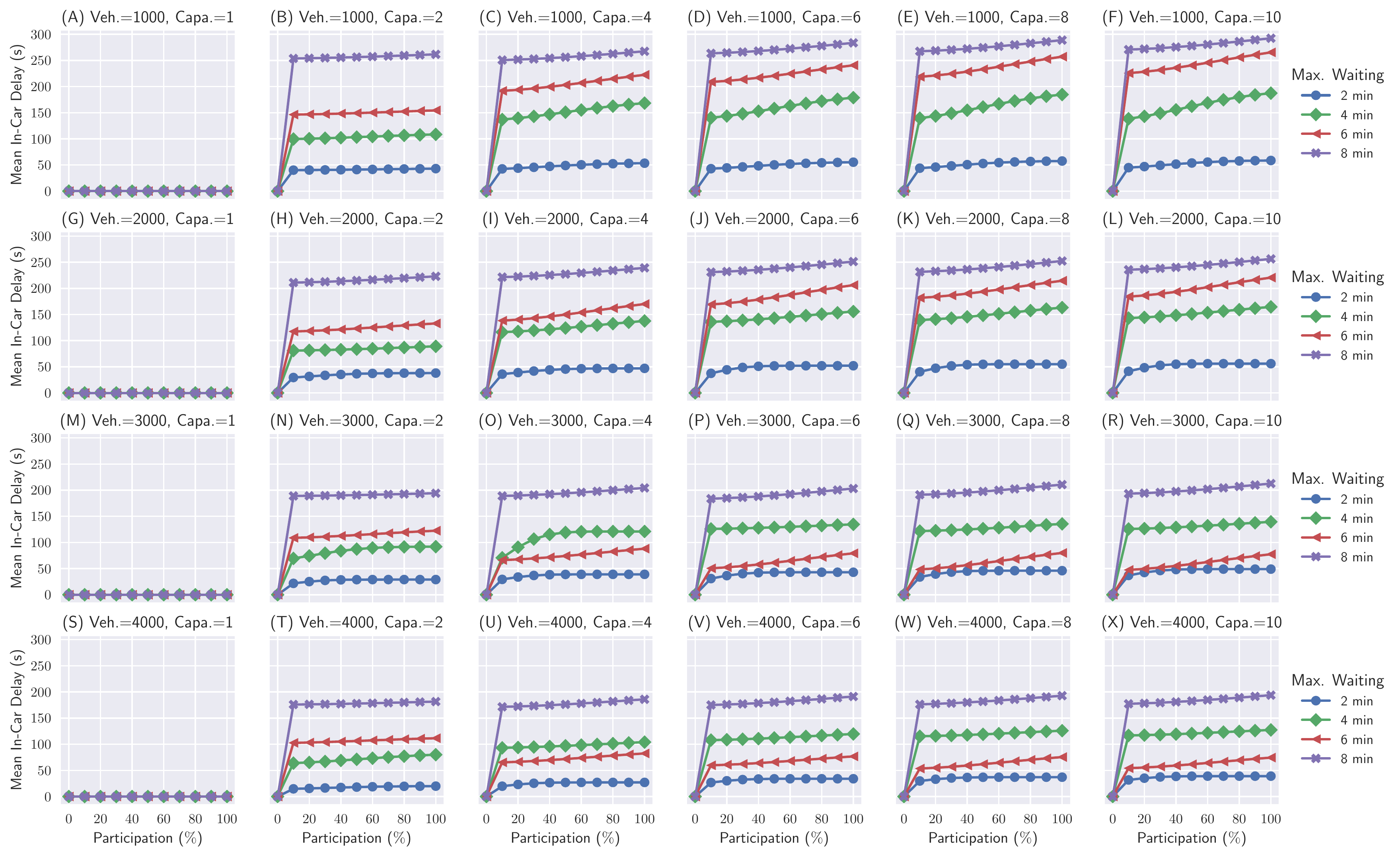}
    \caption{Mean in-car delay time for autonomous ridesharing systems with different configurations.}
      \label{fig:incardelays}
\end{figure*}

\begin{figure*}
\centering
\includegraphics[width=\linewidth]{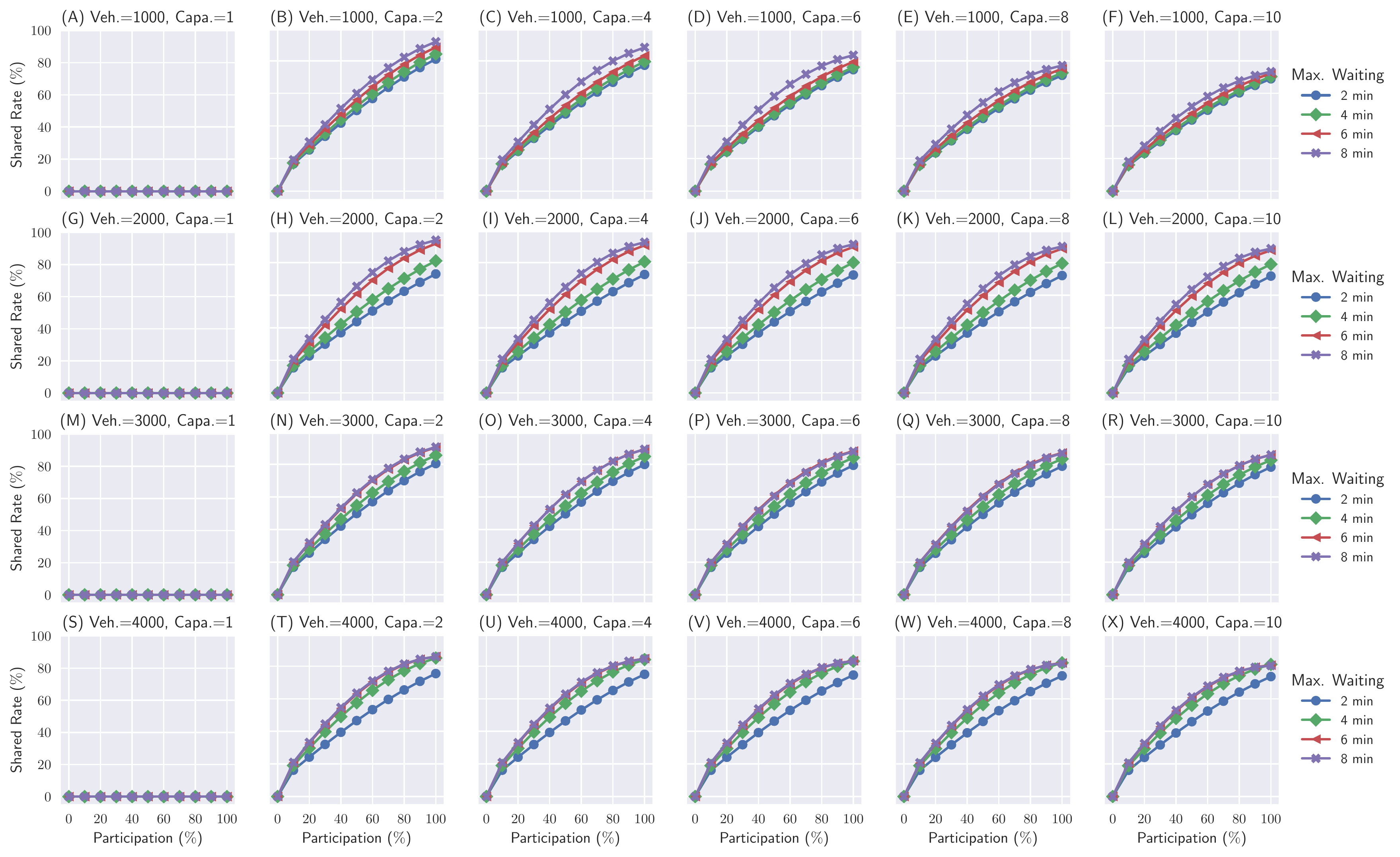}
    \caption{Percentage of shared rides for autonomous ridesharing systems with different configurations.}
      \label{fig:sharedride}
\end{figure*}

\begin{figure*}
\centering
\includegraphics[width=\linewidth]{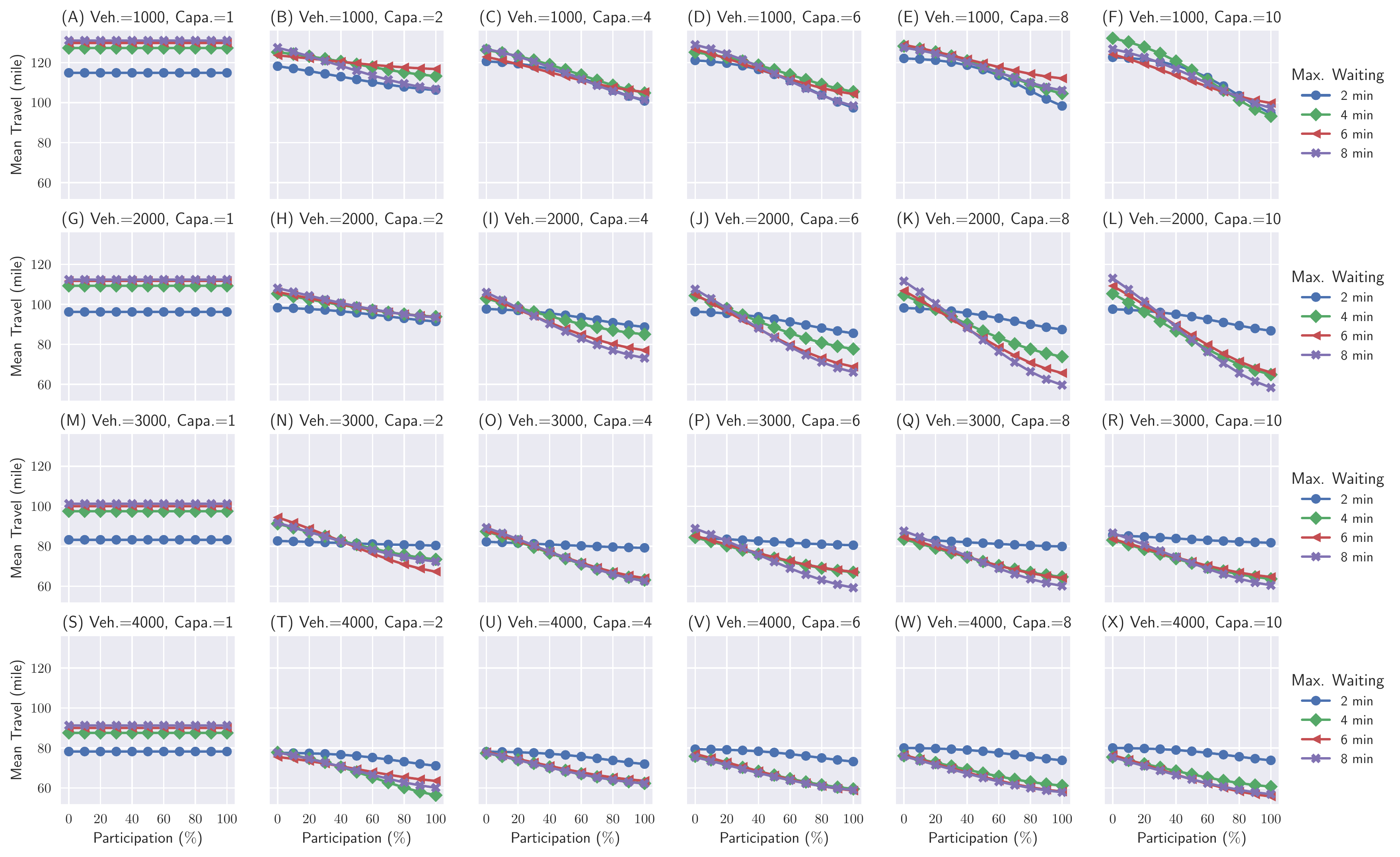}
    \caption{Average distance traveled by each vehicle during a single day for autonomous ridesharing systems with different configurations.}
      \label{fig:trdistance}
\end{figure*}
\end{document}